\documentclass[prd,preprint,nofootinbib,preprintnumbers,longbibliography,superscriptaddress,tightenlines]{revtex4-1}

\usepackage{amsmath,mathrsfs}    
\usepackage{graphicx}   
\usepackage[caption=false]{subfig}
\usepackage{color}
\usepackage[colorlinks=true,linkcolor=blue,citecolor=blue,urlcolor=blue]{hyperref}   
\usepackage{bm} 
\usepackage[inline]{enumitem}

\usepackage{graphicx}
\usepackage{amsmath,amssymb}
\usepackage{textcomp}
\usepackage{hyperref}
\usepackage[capitalise]{cleveref}
\usepackage[toc,page]{appendix}
\usepackage{xcolor}
\usepackage{tikz}

\def\st{\begin{equation}}
\def\stp{\end{equation}}

\usepackage{array}   
\newcolumntype{C}{>{$}c<{$}} 

\definecolor{MyDarkBlue}{RGB}{22, 79, 134}
\definecolor{MyLightRed}{RGB}{200, 37, 6}
\def\half{{\textstyle\frac{1}{2}}}

\def\x{{\bm x}}

\def\k{{\bm k}}

\def\k{{\bm k}}
\def\F{{\mathcal H}}

\def\st{\begin{equation}}
\def\stp{\end{equation}}
\def\bg{\begin{eqnarray}}
\def\nd{\end{eqnarray}}

\def\llangle{\left\langle}
\def\rrangle{\right\rangle}
\def \bes {\begin{subequations}}
\def \ees {\end{subequations}}

\def \F

\def\chemconst{{\chizero}}

\newcommand{\chizero}{{\chi_I}}

\newcommand{\tred}{{\sf t}_{\sf r} }
\newcommand{\tredini}{{\sf t}_{\sf r}^0}
\newcommand{\sigmaeq}{\bar\sigma_{\rm eq}}

\newcommand{\nhat}{\hat{n}}

\def\bsigma{\bar{\sigma}_{\rm eq}}

\newcommand{\DT}[1]{#1}

\newcommand{\dd}{\mathrm{d}}

\newcommand{\mGamma}{\Gamma_0} 

\usepackage{tikz}
\usepackage{pgfplots}
\usepackage{ifthen}

\begin{document}

\title{Quenching through the QCD chiral phase transition}
\author{Adrien Florio}
\email[]{aflorio@physik.uni-bielefeld.de}
\affiliation{Fakultät für Physik, Universität Bielefeld, D-33615 Bielefeld, Germany}
\affiliation{Department of Physics, Brookhaven National Laboratory, Upton, New
York 11973-5000, USA}
\author{Eduardo Grossi}
\email[]{eduardo.grossi@unifi.it}
\affiliation{Dipartimento di Fisica, Universit\`a di Firenze and INFN Sezione
    di Firenze, via G. Sansone 1,
50019 Sesto Fiorentino, Italy}
\author{Aleksas Mazeliauskas}
\email[]{a.mazeliauskas@thphys.uni-heidelberg.de}
\affiliation{Institut für Theoretische Physik, Universität Heidelberg
D-69120 Heidelberg, Germany}
\author{Alexander Soloviev}
\email[]{alexander.soloviev@fmf.uni-lj.si}
\affiliation{Faculty of Mathematics and Physics, University of Ljubljana,
Jadranska ulica 19, SI-1000 Ljubljana, Slovenia}
\author{Derek Teaney}
\email[]{derek.teaney@stonybrook.edu}
\affiliation{Center for Nuclear Theory, Department of Physics and Astronomy,
Stony Brook University, New York 11794-3800, USA}

\date{\today}
   \begin{abstract}
We present a detailed numerical and analytical study of the out-of-equilibrium
dynamics of Model~G, the dynamical universality class relevant to the chiral
phase transition.
We perform numerical 3D stochastic (Langevin) simulations of the $O(4)$ critical point for large lattices in the chiral limit.
We quench the system from the high-temperature unbroken phase to the broken
phase and study the non-equilibrium dynamics of pion fields.
Strikingly, the non-equilibrium evolution of the two-point functions exhibits a
regime of growth, a parametrically  large enhancement, and a subsequent slow
relaxation to equilibrium.
We analyze our numerical results using dynamic critical
scaling and mean-field theory.  The growth of the two point functions is determined by the non-linear dynamics of an  ideal non-abelian superfluid, which is a limit of Model G that reflects the broken chiral symmetry. 
We also relate the non-equilibrium two-point functions to a long-lived parametric enhancement of soft pion yields relative to thermal equilibrium following a quench. 
\end{abstract}

\maketitle

\clearpage

\tableofcontents

\clearpage

\section{Introduction}

One of the original motivations for high-energy collisions of heavy ions was the creation of an ``abnormal'' state of matter with symmetry
properties different from that of vacuum~\cite{osti_4061527}. A prominent example is the symmetry between left and right-handed quarks in QCD. Chiral symmetry is spontaneously broken in a vacuum, which explains the existence of light pions---the (pseudo-)Goldstone bosons\footnote{Chiral symmetry is also explicitly broken by small,  but non-zero up and down quark masses, which leads to light, but not massless pions.}.
After 50 years of research, there is ample evidence that
a new form of chirally symmetric state of matter---the quark-gluon plasma (QGP)---is created in high-energy nuclear collisions at the Large Hadron Collider (LHC) at CERN
and the Relativistic Heavy Ion Collider (RHIC) at BNL~\cite{Busza:2018rrf}. The heavy-ion models based on the viscous hydrodynamic expansion of the QGP
have shown remarkable success in describing experimental low-momentum hadron data. The sophisticated Bayesian fits have put tight constraints on the bulk properties of QGP matter, such as shear and bulk viscosities~\cite{Nijs:2022rme,JETSCAPE:2020shq}. However, the restoration of chiral symmetry in high-temperature QGP plays a surprisingly small role in heavy-ion phenomenology~\cite{Citron:2018lsq,ALICE:2022wpn,Harris:2023tti}.

Hydrodynamics is the universal description of the time evolution of conserved charges. However, in the vicinity of a second-order phase transition, the order parameter evolves asymptotically slowly due to critical slowing down. As such, ordinary hydrodynamics has to be supplemented by additional slowly evolving degrees of freedom to obtain a complete low-momentum description~\cite{hohenberg}. In other words, the fluid-dynamic theory has to include also the
nonconserved variables that evolve on a time scale similar to the conserved
charges to be consistent.
The resulting effective description of the dynamics depends on the interaction
between the hydrodynamical modes and the order parameter. 
This naturally leads to the question if presently
unaccounted-for critical dynamics in heavy ion collisions might be responsible for soft pion enhancement.

An intriguing indication that the current hydrodynamic models might be incomplete is the observed enhancement of low-momentum pions over theory predictions in PbPb collisions at $\sqrt{s_\text{NN}}=2.76~\text{TeV}$ and $\sqrt{s_\text{NN}}=5.02~\text{TeV}$ as measured by ALICE experiment~\cite{ALICE:2013mez,ALICE:2019hno}. Neither complete hydrodynamic simulations~\cite{Devetak:2019lsk,Nijs:2020roc,JETSCAPE:2020mzn,Lu:2024upk}, nor dedicated blast-wave type fits~\cite{ALICE:2013mez,ALICE:2019hno,Mazeliauskas:2019ifr,Melo:2019mpn} are able to account for $\sim50\%$ enhancement of pions below $p_T<0.5~\text{GeV}$. There have been several attempts to explain the soft pion enhancement using the formation of Bose-Einstein condensation of pions~\cite{Begun:2015ifa}, decays of broad resonances~\cite{Huovinen:2016xxq} or perturbative contributions~\cite{Kanakubo:2022ual}.
In early '90s, it was proposed that a sudden quench from chirally symmetric to a broken phase could lead to the formation of domains of disoriented chiral condensate~\cite{Rajagopal:1993ah,Blaizot:1992at,Bjorken:1993cz}. A characteristic modification of neutral to charged pion ratio was looked into in early heavy-ion data, but never confirmed~\cite{Mohanty:2005mv}. Three decades after these pioneering studies, we are at the advantage of tremendous computational and technological progress to study the full extent of chiral critical dynamics in heavy-ion collisions. 
Thanks to the high energy available at the LHC, the number of produced hadrons can reach tens of thousands in a single lead-lead collision---an order of magnitude increase from pre-2000 experiments~\cite{ALICE:2022wpn}. With the advent of high-precision data from high-luminosity LHC and the next-generation heavy-ion experiment, ALICE3~\cite{ALICE:2022wwr}, it is therefore high-time to study the phenomenological signatures of the chiral phase transition in high-energy nuclear collisions.

Critical dynamics in QCD are relevant in several ways.  At low energy nuclear collisions, critical fluctuations in the vicinity of the conjectured QCD critical point, which is the endpoint of the first-order phase transition line between QGP and hadronic phase, are subject to active experimental and theoretical searches~\cite{Pandav:2022xxx}.
In the Hohenberg and Halperin classification scheme \cite{hohenberg}, the relevant critical theory is ``Model H'' \cite{Son:2004iv}, and has recently been studied in the QCD context in \cite{Chattopadhyay:2024jlh}, see also \cite{Chattopadhyay:2023jfm,Roth:2024hcu} for related studies.
At high-energy collisions at the LHC, the high-density baryon-symmetric matter cools down and transitions from a chirally symmetric deconfined phase to a hadronic phase.
While this transition is not second-order per se, the QGP undergoes a rapid chiral cross-over~\cite{Borsanyi:2018grb,HotQCD:2018pds}. It is therefore reasonable to expect the dynamics of the order parameters to be parametrically slow. In the limiting case of chiral
QCD, when $m_{up}=m_{down}=0$, the phase transition is now known to be second order \cite{HotQCD:2019xnw,Kaczmarek:2020sif,Kotov:2021rah,Cuteri:2021ikv}.
These computations show that even for realistic quark masses, the QCD susceptibilities are well described by universal scaling functions of the $O(4)$ static universality class.
The appropriate description of the associated dynamic universality class is ``Model
G''~\cite{hohenberg}. Model G is an effective  field theory of pions (pion EFT), the Goldstone particles associated with the
 $O(4)\simeq SU(2)_L\times SU(2)_R$ spontaneous symmetry breaking, coupled to the chiral condensate.
The first
 attempt at resolving Model G's dynamics in the context of QCD was presented in
 \cite{Schlichting:2019tbr} and subsequent works have carefully studied
 relevant universal properties and scaling functions \cite{Florio:2021jlx,Florio:2023kmy}.
 
In this work we study, for the first time to our knowledge,  the out-of-equilibrium
dynamics of Model G.
We perform 3D Langevin simulations of an $O(4)$ symmetric scalar field (which is coupled to conserved charges) quenched from the symmetric to the broken phase close to the critical point.
We find that a sudden quench of the system generates a characteristic growth of the order parameter.
This growth is  coupled to a corresponding parametric enhancement in the
(equal time) pion correlation function, which subsequently relaxes slowly back to equilibrium. While this is reminiscent of the situation in many cosmological preheating scenarios~\cite{Amin:2014eta,Boyanovsky:1996sq,Berges:2002cz}, universality and the non-trivial hydrodynamic limit of Model G's pion EFT quantitatively control the structure of the parametric enhancement, making the dynamics novel and markedly different from these studies.  In particular, the growth of the condensate is controlled by the non-linear and non-dissipative dynamics of an $SU(2)_L\times SU(2)_R$ superfluid theory, reflecting the broken chiral symmetry. 
A brief account of this work is given in a companion letter~\cite{Florio:2025zqv}.

In more detail, this paper is structured as follows. In
\cref{sec:modelG}, we review Model G, which we use to investigate the
non-equilibrium dynamics. In particular we discuss the relevant timescales governing the pion dynamics in the broken phase. In  \cref{sec:quenches}, we go fully
non-adiabatic and study the case of an instantaneous quench from the hot to the
cold phase. We first use the scaling analysis, independent of our simulation details, to show the expected parametric behavior of the condensate and two-point correlation functions. We confirm the scaling analysis with lattice simulations of quenches. Finally, we show that the mean-field analysis, which provides a simple analytic model of the quenches, fails in several important aspects.
In \cref{sec:discussion} we summarize our results and present arguments of why the observed pion enhancement will persist with explicit symmetry breaking. 
\Cref{sec:subleading} provides a concise summary of the leading and subleading critical behavior of the equilibrium quantities established in earlier studies.
In \cref{app:meanfield} we provide details on mean-field analysis.

\section{Overview of Model G and its phases}
\label{sec:modelG}

 As already discussed in the introduction, critical dynamics is the appropriate framework
 to discuss the time evolution near the chiral phase transition 
 of QCD.
 The relevant 
 dynamical critical model is
 Model G~\cite{Rajagopal:1992qz,Grossi:2020ezz,Grossi:2021gqi} and describes the universal dynamics of pions, the Goldstone particles associated to the $O(4)\simeq SU(2)_L\times SU(2)_R$ spontaneous symmetry breaking. 

The static properties of QCD near the chiral transition can be
modeled with an $O(4)$ order parameter
$\phi_a=(\phi_0,\phi_1,\phi_2,\phi_3)$,  which parametrizes the quark
condensate\footnote{As in our previous work $O(4)$ indices are indicated
    by $a, b, c=0\ldots 3$~\cite{Grossi:2021gqi,Florio:2021jlx}. The ``spatial'' components describing the $O(3)$
isospin subgroup are indicated with an arrow, e.g.  $\phi_a = (\phi_0,
\vec{\pi})$.  If necessary,  an index $\ell$ labels the components,
$\vec{\pi}_\ell=  (\phi_1,  \phi_2, \phi_3)$. The dot product indicates an
appropriate contraction of indices when clear from context, e.g. $\phi
\cdot \phi = \phi_a \phi_a$, $\vec{\pi} \cdot \vec{\pi} = \vec{\pi}_\ell \vec{\pi}_\ell$.},  $\llangle \bar{q}_R q_L \rrangle$.  
The corresponding Landau-Ginzburg free energy of the $O(4)$ model is 
\begin{equation}
    \label{eq:Landauginzburg}
    \mathcal H[\phi]  = \int d^dx \left[\frac{1}{2} \nabla \phi_a \cdot 
\nabla \phi_a + U(\phi) - H \cdot \phi  \right]\, ,
\end{equation}
where $d=3$ is the number of spatial dimensions, and the potential $U(\phi)$ is 
\begin{equation}
    \label{eq:potential}
U(\phi) = \frac{1}{2} m_0^2 \, \phi^2  + \frac{\lambda}{4} (\phi\cdot\phi)^2 \,
,
\end{equation}
with $m_0^2$ negative. 
The magnetic field  $H_a= (H,\vec{0})$ in the model accounts for the explicit breaking of $O(4)$ symmetry by the quark mass in QCD. 
In equilibrium, the order parameter is distributed according to the statistical weight:
\st
P[\phi ]  = \frac{1}{Z} e^{-\mathcal H[\phi]/T_c} \, .
\stp

The $O(4)$ model with $\lambda=4$ and $H=0$ has a second order phase transition at $m_0^2=m_c^2=-4.8110(4)$ \cite{Florio:2021jlx}, where the order parameter acquires a non-zero expectation value. 
All models belonging to the $O(4)$ universality class, including QCD and the model in \cref{eq:Landauginzburg}, will exhibit the same critical behavior as a function of relevant couplings, which are normalized by convention.
 In the chiral limit ($H=0$), the model in \cref{eq:Landauginzburg} has one relevant coupling  $m_0^2$ which can be varied around $m_c^2$. 
We define the reduced temperature in the simulation, $\tred$, as 
the deviation of the relevant coupling from its critical value\footnote{In mean field theory, $m_c^2=0$ and the reduced temperature is defined as
\begin{align}
    \tred = \frac{m_0^2}{\mathfrak{m}^2},
\end{align}
where $\mathfrak{m}$ is a dimensionful parameter, scaling as the microscopic length to the power of $-1$.}

\st
\tred \equiv  \frac{m_0^2 - m_c^2 }{|m_c^2|} \, . 
\label{eq:trOne}
\stp
Previously in~\cite{Florio:2021jlx} we matched the reduced temperature and field of \cref{eq:Landauginzburg}  to a conventional parametrization of the $O(4)$ magnetic equation of state given in \cite{Engels:2014bra}. 
For QCD in the chiral limit,  the only relevant coupling close for chiral phase transition is the temperature, and the reduced temperature is $(T - T_c)/T_c$. 
Ultimately the relation between $\tred$ and $(T-T_c)/T_c$ in QCD  must be determined by matching lattice QCD data on the chiral condensate to the conventional parametrization discussed above~\cite{Kaczmarek:2020sif}.

The real time  critical dynamics of a second order phase transition depend on the order parameter and the number and type of conserved charges~\cite{hohenberg}. 
In the case of the QCD and the  chiral phase
transition, both the iso-vector 
charge,  $\vec{n}_V\sim \bar q \gamma^0 \vec{t}_I q$, 
and iso-axial vector charge,  $\vec{n}_A\sim \bar q \gamma^0 \gamma^5\vec{t}_I q$, are conserved and couple to the order parameter. In the $O(4)$ language, $(\vec{n}_V, \vec{n}_A)$ are combined into an antisymmetric $O(4)$ tensor $n_{ab}$ with 
\begin{align}
    n_{0\ell} = n_A^\ell\,,   \qquad   n_V^{\ell} =\frac{1}{2} \epsilon^{\ell\ell_1\ell_2 }  
    n_{\ell_1\ell_2}.   
\end{align}
As these conserved charges are non-critical, their equilibrium distribution is Gaussian
\begin{equation}
    \mathcal H_n \equiv  \int d^3x \left[\frac{1}{4\chizero} n_{ab}n^{ab}  \right]\,
,
\label{eq:chargesusc}
\end{equation}
where $\chizero$ is the isospin susceptibility\footnote{The isospin susceptibility $\chizero$ is unrelated to the susceptibility of the order parameter $\chi$ defined in \cref{eq:chi}.}.

The interactions of the conserved charges with the order parameter arise from nontrivial Poisson brackets --- see \cite{Rajagopal:1992qz,Grossi:2021gqi} for the expressions and explicit derivation. The effective equations of motion that describe the system in the critical region, close to equilibrium, are
\begin{subequations}
    \label{eq:eom}
\begin{align}
   \partial_t \phi_a  + \frac{g_0}{\chizero}\,n_{ab} \phi_b  
  &=\Gamma_0 \nabla^2 \phi_a - \Gamma_0 (m_0^2 + \lambda \phi^2)\phi_a +
   \Gamma_0 H_a + \theta_a\, ,\label{eq:eom1}\\
   \partial_t n_{ab}  + g_0 \,\nabla \cdot (\nabla \phi_{[a} \phi_{b]}) + g_0 H_{[a}
   \phi_{b]} 
             &= D_0 \nabla^2 n_{ab} +  \partial_{i} \Xi_{ab}^i \label{eq:eom2}\, .
\end{align}
\end{subequations}
Here  $H_{[a}\phi_{b]}$ denotes the anti-symmetrization, $H_a \phi_b - H_b
\phi_a$. The coefficients  $\Gamma_0$ and $D_0$  are the bare kinetic
coefficients associated with the order parameter and the charges\footnote{Outside of $d=4$ (mean field), the renormalized parameters $\Gamma$ and $D$ are markedly different from the bare ones, and are defined by a pion EFT described in \cref{sec:hydroGM}.}. The bare
conductivity of the charges is related to the bare diffusion coefficient and the charge susceptibility,  $\sigma_0 =  \chizero D_0$. The constant
$g_0$ is a coupling of the field $\phi$, and has the units of $({\rm
action})^{-1}$ in our conventions.
Finally, $\theta_a$ and $\Xi_{ab}$ are the appropriate  noises, which  are
defined through their two-point correlations~\cite{hohenberg}
\begin{subequations}
\begin{align}
   \langle \theta_a(t,x)\theta_b(t',x') \rangle &= 2 T_c\Gamma_0 \, \delta_{ab}
   \, \delta(t-t')\delta^3(x-x') \, ,\label{eq:langevin_var}\\
   \langle \Xi^i_{ab}(t,x)\Xi^j_{cd}(t',x') \rangle &= 2  T_c \chizero D_0 \,
   \delta^{ij} \left(\delta_{ac} \delta_{bd} - \delta_{ad} \delta_{bc} \right)
   \, \delta(t-t')\delta^3(x-x') . \label{eq:langevin_var_cons}
\end{align}
\end{subequations}

The dynamical equations listed above  define Model G in the Halperin and Hohenberg classification scheme. The field will equilibrate to the free energy distribution,   $\mathcal H[\phi, n]    \equiv \mathcal H_\phi[\phi] + \mathcal H_n[n] $~\cite{hohenberg}.  
Following our previous work we have discretized the equations of motion on a lattice  of size $L$, setting the lattice spacing $a$, the critical temperature $T_c$, and coupling $g_0$ to unity $g_0=T_c=a=1$~\cite{Florio:2021jlx,Florio:2023kmy}.  \DT{In these units isospin susceptibility and kinetic coefficients are taken to be $\chizero =5$, $\Gamma_0=1$ and $D_0=1/3$~\cite{Florio:2021jlx}}.

\DT{
Our numerical scheme is detailed in Appendix A of \cite{Florio:2021jlx}. Briefly, the scheme uses operator splitting, first evolving the  system with the ideal (Poisson-bracket) terms, using a second order position-Verlet symplectic integrator. The time step is $\Delta t = 0.24 \,a^2/\Gamma_0$ . Then, six dissipative steps (each of step size $0.04 \, a^2/\Gamma_0$) are taken using a Metropolis scheme. In this scheme,  a proposal is made for  the field update at a given lattice site,  $\phi_a(\x) \rightarrow \phi_a(\x) + \delta\phi_a(\x) $,  with variance $\langle \delta \phi_a^2  \rangle_0 = 2 T_c\Gamma_0 \Delta t / a^3 $. This proposal is then accepted or rejected with probability,  ${\rm min}(1, \exp(-\Delta \mathcal H/ T_c ))$, where $\Delta {\mathcal H} = \mathcal{H}[\phi_a + \delta \phi_a] - \mathcal{H}[\phi_a]$. The accept-reject process introduces the dissipation  and reproduces the Fluctuation Dissipation Theorem (FDT). The dissipative time step,  $0.04 \, a^2/\Gamma_0$,  is chosen so that the accept-reject probability is approximately one half. A similar Metropolis step is also used to update the charges, reproducing the stochastic diffusion equation. Full details are given in Appendix A  of \cite{Florio:2021jlx}. }

As summarized below, we have previously analyzed the real-time equilibrium correlation functions using this scheme, both above and below  $T_c$~\cite{Florio:2021jlx,Florio:2023kmy}. 

\subsection{Equilibrium dynamics of the order parameter above $T_c$}

In this section we will briefly review the  dynamics of the order parameters in equilibrium and above  $T_c$~\cite{tauber2014critical,Goldenfeld:1992qy}.
In this regime
the correlation length $\xi$ sets the relevant length scale,  and perturbations of the order parameter relax slowly back to equilibrium on a dynamical time scale,  $\tau_R \propto \xi^{\zeta}$,  where $\zeta=d/2$ is the dynamical critical exponent of Model G. Although we will speak informally about these scales, they can be precisely defined from specific equilibrium correlators.  

The magnetization is defined from the zero mode of the order parameter
\begin{equation}
    \label{eq:Mdef}
    M_a(t) \equiv \left. \phi_a(t, {\bm k})   \right|_{\k= 0} \,, \qquad   \phi_a(t, \k) \equiv \frac{1}{V} \sum_{\bm x} e^{-i\k \cdot {\bm x}} \phi_a(t, {\bm x}),
\end{equation}
where $V\equiv L^d$ is the three-dimensional lattice volume for $d=3$.
The magnetic susceptibility $\chi$ at  $H=0$ scales as
\begin{equation}
    \label{eq:chi} \chi \equiv \frac{V}{4} \sum_{a=1}^4 \left \langle M_a(t) M_a(t) \right \rangle_{\rm eq} = C^+
    \tred^{-\gamma} \sim \xi^{2-\eta} \, . 
\end{equation}
Here the average $\llangle \ldots \rrangle_{\rm eq}$  
denotes an average over time, which is equivalent to the ensemble average.
$\chi\equiv \chi(\tred)$ is a function of reduced temperature, but here and below we will drop the $\tred$ argument when confusion does not arise.  
The critical exponents $\gamma$ and $\eta$ are given in \cref{tab:num_values_fixed}. 
More generally at finite $\k$ we define the static correlation function
\begin{equation}  
    G_{\phi\phi}(k) \equiv \frac{V}{4}\sum_{a=1}^{4}\llangle |\phi_a(t, {\bm k})|^2 \rrangle_{\rm eq} \, , 
\end{equation}
and we will implicitly average over the three Cartesian directions for  a specified $k=|\k|$ to increase statistics.
When the correlation length must be precisely defined, for $T>T_c$ we use the standard ``second moment'' correlation length\footnote{The motivation for the definition is the following -- in continuous mean field theory $G_{\phi\phi}(k) = \chi/((k\xi)^2 + 1)$. Now replace $k^2$ in the mean field formula with the lowest eigenvalue of
the discretized Laplacian,  $k^2 \rightarrow (2\sin(ka/2)/a)^2$,  with lattice spacing $a=1$. Solving for $\xi$ produces \cref{eq:xidef}.  } used in \cite{Hasenbusch:2021rse,Florio:2023kmy} : 
\begin{equation}
    \xi \equiv \frac{1}{2\sin(\pi/L)}
    \sqrt{\frac{ \chi }{\left.G_{\phi\phi}(k)\right|_{k=2\pi/L}}- 1} = \xi^+ \tred^{-\nu} \qquad  \mbox{for $\tred > 0$ }
           \, .  \label{eq:xidef}  
\end{equation} 
Below $T_c$ we will use a different definition of the correlation length based on the pion decay constant as discussed below. 

References \cite{Florio:2021jlx,Florio:2023kmy} provide a careful study of the critical behavior of this model, including the determination of relevant non-universal constants $C^+$ and $\xi^+$  and the first correction to scaling -- we refer the reader to \cref{tab:subleading} and the Appendix B of \cite{Florio:2023kmy} for the complete parameterization of the equilibrium properties of the model obtained from Langevin dynamics of the previous section.

The order parameter relaxation time can be defined from real time response of the system.  For instance,  previously we analyzed the unequal time correlation function 
\st
\label{eq:Gphiphiresponse}
G_{\phi\phi} (\Delta t, k) \equiv  \frac{V}{4 } \sum_a \llangle \phi_a(t + \Delta t, \k ) \phi_a(t, -\k) \rrangle_{\rm eq} \,  , 
\stp
and, motivated by exponential decay,  defined a relaxation time based on the  integral 
\st
\tau_R   \equiv  \chi^{-1} \int_0^{\infty} \dd \Delta t \,  G_{\phi\phi}^{\rm eq} (\Delta t, 0)   \, .
\stp
The relaxation time diverges in the vicinity of the critical point in a universal fashion 
\begin{equation}
    \tau_R = \tau^+ \tred^{-\nu \zeta}  \sim  \xi^{\zeta} \, , 
\end{equation}
where $\zeta=d/2$ is the dynamical critical exponent of Model G and $\tau^+$ is a microscopic timescale. Again we previously measured the non-universal constant $\tau^+$ (see \cref{tab:subleading}) and demonstrated the dynamical scaling of the model in a number of ways~\cite{Florio:2021jlx,Florio:2023kmy}. 

We note that all of the scaling relations above follow from the scaling hypothesis, where the correlation function takes the scaling for~\cite{tauber2014critical,Goldenfeld:1992qy}
\st
G_{\phi\phi}( \Delta t, k, \tred) = s^{2 - \eta}  G_{\phi\phi}( s^{-\zeta} \Delta t, s k, s^{1/\nu} \tred) \, , 
 \stp
 with  $s$ a positive number. 
For instance,  setting $k=0$,  $\Delta t=0$ and $s=\tred^{-\nu}$ reproduces the temperature dependence of the susceptibility $\chi \propto \tred^{-\gamma}$,  after noting the relation $\gamma = \nu(2 - \eta)$ given in \cref{tab:num_values_fixed}. 

\setlength{\tabcolsep}{10pt} 
\renewcommand{\arraystretch}{1.5}
\begin{table}
  \centering
  \begin{tabular}{|c|c|c| c | } 
  \hline
  quantity & scaling & relation  &  exponent value   \\
  \hline
  correlation length & $\xi \propto \tred^{-\nu}$ &  & $\nu =0.7377(41)$   \\
susceptibility &  $\chi \sim \xi^{2 - \eta} \propto \tred^{-\gamma} $ & $\gamma = \nu (2 -\eta)$ & $\eta = 0.0302(16)$   \\
  condensate &  $\bar{\sigma}_{\rm eq}
  \sim \xi^{-(d-2 + \eta)/2}
  \propto (-\tred)^{\beta} $
             &  $\beta = \nu (d - 2 + \eta)/2 $ & 
  $\beta=0.380(2)$  \\ 
relaxation time & $ \tau_R \propto \xi^{\zeta}$ &  &  $\zeta= d/2$\\
  \hline
\end{tabular}
\caption{Critical exponents of the $O(4)$ model at zero magnetic field used in this work.
    The high precision values for the static exponents are from Engels {\it et al.}~\cite{Engels:2011km,Engels:2014bra}. Within the uncertainty of our simulation,  these exponents are reproduced by the Langevin simulations after accounting for subleading corrections~\cite{Florio:2021jlx,Florio:2023kmy}. The dynamical critical exponent $\zeta$ is also reproduced by these simulations.
}
\label{tab:num_values_fixed}
\end{table}

\DT{
\subsection{Equilibrium dynamics of Goldstone modes below $T_c$ }

In this section, we will briefly review the equilibrium dynamics of pions 
below $T_c$ following Son and Stephanov and our previous work~\cite{Son:2001ff,Son:2002ci,Grossi:2021gqi}.   First we provide an overview of pion propagation close to the critical point.   Subsequently we review the hydrodynamic equations of motion that lead to these waves. Finally,  we connect these classical waves to a quasi-particle description.
}
\subsubsection{Overview}
\label{sec:pionoverview}
Below $T_c$, the system spontaneously breaks the $O(4)$ symmetry, thereby developing a condensate.  The typical correlation length below $T_c$ remains of order $\xi \sim |\tred|^{-\nu}$ and the typical relaxation time for modes with $k \xi \sim 1$ remains of order $\tau_R \sim \xi^\zeta$ with  $\zeta=d/2$. 
However,  long wavelength pion modes of length $L \gg \xi$  are characterized by the longer time scales $\tau_R\, (L/\xi)$ and $\tau_R\, (L/\xi)^2$.  Analyzing the statics and dynamics of long wavelength pions is the best way to precisely define the parameters $\xi$ and $\tau_R$ below $T_c$.  

Specifically, by analyzing the spatial correlation function of pions at large distances, one determines the spatial decay constant, $f^2$,  which scales as $\xi^{2-d}$ (see \cite{Son:2001ff} and the discussion around \cref{eq:Gpipieq}).  Thus in three dimensions $f^2$
provides a definition of the correlation length,   $\xi  \propto 1/f^2$.  Similarly the pion dispersion curve at small wave numbers takes the form 
\st
\label{eq:dispersion}
    \omega(k) = v k - \frac{i}{2} D_A k^2 \, , 
\stp
Here the velocity and diffusion parameters in the simulation,  $v$ and $D_A$,  depend critically on the reduced temperature and correlation length. The parameters can be (and were!) precisely determined by analyzing pion waves close to the critical point~\cite{Florio:2023kmy}.  
As discussed below, the pion velocity is defined as 
\st \label{eq:velocity}
v \equiv \sqrt{\frac{f^2}{\chizero}} \, , 
\stp
where $\chizero$ is the charge susceptibility  of \cref{eq:chargesusc} and is constant near the critical point.  Thus the velocity 
scales as 
\st
v \sim \xi^{1- d/2}  \sim \frac{\xi}{\tau_R} \, , 
\stp
close to the critical point~\cite{Son:2001ff}. \DT{See \cref{tab:subleading} and \cite{Florio:2021jlx,Florio:2023kmy} for a parametrization including subleading corrections.}
The ballistic propagation of pions over length $L$ takes a time of order,  $L/v \sim  \tau_R\, (L/\xi)$.  

The damping coefficient $D_A$ must scale as $\xi^2/\tau_R$ so that when $k$ is of order $\xi^{-1}$,   the frequency $\omega(k)$ is of order $\omega \sim 1/\tau_R$, in accord with the scaling hypothesis. Note that for $k\xi\sim 1$ the real and imaginary parts of $\omega$ are the same order of magnitude~\cite{Son:2001ff,Son:2002ci}. Thus,  the decay time for the pion mode of size  $L \gg \xi$ is of order $\tau_R (L/\xi)^2$.   

To summarize, soft pions are characterized by their spatial decay constant $f^2 \propto 1/\xi$.  For pions of wavelength $L \gg \xi$, the  time scales associated with  ballistic pion propagation and diffusion are:
\st
  \mathrm{ballistic\ time} \equiv \frac{L}{v}  \sim   \tau_R \left(\frac{L}{\xi}\right)   \, , 
 \quad \mbox{and} \quad~ \mathrm{diffusion\ time}  \equiv \frac{L^2}{D_A} \sim \tau_R \left(\frac{L}{\xi} \right)^2 \, .
 \label{eq:timescales}
\stp
The pion description is valid for $k\xi \ll 1$ and times  $t \gg \tau_R$. We will now elaborate on the equilibrium pion effective theory, before studying the non-equilibrium dynamics.

\subsubsection{Spontaneous symmetry breaking}

Below the critical temperature the system spontaneously condenses. The equilibrium value of the condensate is defined from the limiting procedure~\cite{Hasenfratz:1989pk} 
\st
\bar{\sigma}_{\rm eq} \equiv \lim_{H\rightarrow 0} \lim_{V \rightarrow \infty }  \llangle M_0 \rrangle_{\rm eq} \, ,
\stp
where $V$ is  taken to infinity first,  and then $H$ is taken to zero. 
The mean magnetization drops towards the critical point as 
\st
\bar \sigma_{\rm eq} = B^- (-\tred)^{\beta} \, , 
 \stp
 where $\beta$ is given in \cref{tab:num_values_fixed} and the non-universal constant $B^-$ and subleading correction were determined in our previous work~\cite{Florio:2021jlx}, see \cref{tab:subleading}. 

Even at zero magnetic field and finite volume the condensate can still be determined from the correlations in the system.  The zero mode of the order parameter is decomposed into  a magnitude $\sigma(t)$ and a direction  $\nhat_a(t)$
that generally depends on time
\st
 M_a(t) \equiv  \sigma(t) \nhat_a(t) \, . 
\stp
However, in large enough volume and below the critical point   the value of $\sigma$ and the direction $\hat n_a$ are nearly constant in time\footnote{
    More precisely $\nhat_a(t)$ diffuses over the three sphere on a timescale proportional to  $L^{3}$ with $\tau_R$ and $\xi$ making up the remaining dimensions, $\sim \tau_R (L/\xi)^3$~\cite{Florio:2023kmy}.
The equilibration time of the system is proportional to $\tau_R \, (L/\xi)^2$ and is determined by the damping rate of Goldstone modes (see the main text).
Thus, starting from a random initial condition,  the mean magnetization can be  defined as  infinite volume limit of a time average,   with time in a restricted range:
\st
\bar \sigma_{\rm eq} \nhat_a  \equiv \lim_{L\rightarrow \infty} \llangle M_a \rrangle_{\rm eq}       \quad \mbox{with} \quad   {\tau_R} \left(\frac{L}{\xi} \right)^2  \ll t  \ll   \tau_R\,\left( \frac{L}{\xi} \right)^3  \, .
\stp
As discussed around \cref{eq:Hasenfratz}, there are easier ways to determine the condensate numerically.
}.
Using $\nhat_a(t)$ defined from the zero mode, and an additional three orthogonal basis vectors $e^{\ell}_a(t)$ with $\ell = 1\ldots 3$ defined as follows
\st
{\vphantom{e^{\ell}_a}\nhat}_a e^{\ell}_a = 0 \, ,    \qquad   e^{\ell_1}_a  e^{\ell_2}_a = \delta^{\ell_1 \ell_2}   \, ,  \qquad \left(\vec{e}_a \times \vec{e}_b\right)^{\ell_1} \equiv   \varepsilon^{\ell_1 \ell_2 \ell_3} \; e^{\ell_2}_a e^{\ell_3}_b \,  ,  
\stp
the fields can be decomposed as  
\st
\phi_a(t, {\bm x}) \equiv   \sigma(t, {\bm x})  \, \nhat_a(t) + 
\vec{\pi}(t, {\bm x}) \cdot \vec{e}_a(t) \, .
\stp
Here $\vec{\pi}\cdot\vec{e}_a= \pi^\ell \, e^{\ell}_a$  and from its definition,  $\pi^\ell(t, \k=0) = 0$.  
The equal time correlation functions we will need are 
\begin{subequations}
    \label{eq:Gequaltime}
\begin{align}
    G_{\sigma\sigma}(k ) &\equiv   V \llangle \sigma(t, \k) \sigma(t, {-\bm k}) \rrangle_{\rm eq} \, ,\label{eq:Gsgsg} \\
    G_{\pi\pi}(k) &\equiv   \frac{V}{3} \sum_{\ell=1}^{3} \llangle \pi^\ell(t, \k) \pi^\ell(t, {-\bm k}) \rrangle_{\rm eq}  \label{eq:Gpipi} \, ,  
\end{align}
\end{subequations}
The unequal time correlation functions such as $G_{\pi\pi}(\Delta t, k)$ are defined analogously to \cref{eq:Gphiphiresponse}.
The conserved charges are decomposed as
\st
n_{ab}(t, {\bm x})  =
\vec{n}_{A}(t, {\bm x})  \cdot \left( \nhat_a \, \vec{e}_b -  \nhat_b \, \vec{e}_a \right) 
+ 
\vec{n}_{V}(t, {\bm x})  \cdot  \left(\vec{e}_a \times \vec{e}_b \right)
\stp
where as above,  $\vec{n}_V \cdot \vec{e}_a  \equiv n_V^\ell e^{\ell}_a$. 
We will not study the correlations of the charges in this work, but refer 
to our previous work for a detailed study~\cite{Florio:2023kmy}.

\subsubsection{Superfluid hydrodynamics of Goldstone modes}\label{sec:hydroGM}

Next, we will review how the equilibrium parameters of the pion dispersion curve in \cref{eq:dispersion} were extracted from simulations and their theoretical underpinnings. 
For wavelengths longer than the correlation length $ \xi  \ll k^{-1}  $, 
the magnitude of the order parameter $\phi_a(t,{\bm x})$ is approximately constant,  $\sqrt{\phi_a(t,\x) \phi_a(t,\x)} \simeq \bar{\sigma}_{\rm eq}$.  Fluctuations of the order parameter 
are dominated by the changes in the local direction of $\phi_a(t,\x)$ indicated with a ``spin'' vector $s_a(t, {\bm x})$,  which is normalized to unity $s_a s_a=1$ and parametrizes the  Goldstone modes.

For definiteness,  take $\nhat_a=(1, 0,0, 0)$ as the condensate direction. Small fluctuations in the order parameter can be parametrized by three angles $\vec{\varphi}(t,\x)$ indicating the deviation of $s_a(t,\x)$ from $\nhat_a$:  
\st
\phi_a(t, {\bm x}) = (\sigma(t,\x),  \vec{\pi}(t,{\bm x})) =  \bsigma s_a(t, {\bm x}) \, , 
\stp
with
\st
s_a \simeq \left(1 - \tfrac{1}{2} {\vec{\varphi}}^2(t,\x), \vec{\varphi}(t,{\bm x})\right) \, . 
\stp
Here we defined local angles $\vec{\pi} \simeq \bsigma \vec{\varphi}$  from the pion field, and made the small angle approximations,  $\cos\varphi \simeq 1 - \half \vec{\varphi}^2$ and $\vec\varphi \sin(\varphi)/\varphi \simeq \vec{\varphi}$. 
The appropriate effective theory for the Goldstone modes (akin to chiral perturbation theory at zero temperature) is based on parametrizing the fluctuations in spins or angles, and may be recognized as the $SU(2)_L\times SU(2)_R$ generalization of a $U(1)$ superfluid~\cite{Son:1999pa}. 
For the applications discussed in this work (e.g. \cref{fig:drawing1}), the small angle approximation is not enough, but it is sufficient to analyze the  pion dispersion curve discussed previously. 

The free energy associated with the Goldstone modes and charge fluctuations  for $k \ll \xi^{-1}$ is~\cite{Hasenfratz:1989pk} 
\st
\label{eq:spinfree}
\mathcal  H[s_a,n] =\int {\rm d}^dx \, 
\frac{f^2}{2}    \,  \partial_i s_a \cdot  \partial^i s_a  
+ \frac{n_{ab}^2}{4\chizero}  \, ,      
\stp
or for small fluctuations
\st
\label{eq:varphifree}
 \mathcal  H[\varphi,n] \simeq \int {\rm d}^dx \,
     \frac{f^2}{2}    \,  \partial_i \vec{\varphi} \cdot  \partial^i \vec{\varphi}  + 
\frac{\vec{n}_V^2}{2\chizero} + \frac{\vec{n}_A^2}{2\chizero} \, .
\stp
Here $\chizero$ is the charge susceptibility appearing in \cref{eq:chargesusc} and is constant near the critical point\footnote{This is the charge susceptibility $\chizero$ and not the magnetic susceptibility in \cref{eq:chi}.}.
However,  $f^2$, the spatial pion decay constant, is a matching coefficient describing the pion effective theory  and is determined
by integrating out modes which fluctuate on length scales of order the correlation length $\xi$, leaving only the Goldstone modes in the effective theory. Thus, $f^2$ depends critically on the reduced temperature.  

Specifically, near the critical point,  the pion decay constant scales as $f^2 \propto \xi^{2- d }$ as we now review~\cite{Hasenfratz:1989pk,Son:2001ff}. 
Using the Gaussian effective theory in \cref{eq:varphifree}, the  static correlations of the order parameter at long wavelength can be straightforwardly computed\footnote{\DT{Note that $T_c$ has been set to unity. If the temperature is restored,  this equation is written, $G_{\pi\pi}^{\rm eq} = T_c \, \bsigma^2/f^2 k^2$. The factor of temperature is needed when comparing the $G_{\pi\pi}$ to the pion Bose-Einstein distribution below.}} :
\st
G_{\pi\pi}(k) \simeq \bsigma^2 G_{\varphi\varphi}( k) =\frac{\bsigma^2}{f^2 k^2} \sim \frac{\xi^{2- \eta}}{(k \xi)^2 } \,. \label{eq:Gpipieq}
\stp
For $k\xi \ll 1$,  the static pion correlator is enhanced by a factor of $1/(k\xi)^2$ relative to the static sigma correlator, which scales as the magnetic susceptibility $\chi \sim \xi^{2-\eta}$. As $k\xi \rightarrow 1$,  the equilibrium pion and sigma fluctuations must become comparable  to ensure that the system ultimately realizes chiral symmetry restoration.   This requires $G_{\pi\pi}$  to approach  $\xi^{2-\eta}$ in this limit, thereby justifying the scaling $f^2\propto \xi^{2-d}$ a posteriori~\cite{Son:2001ff}.

The value of $\bsigma$ can be estimated from the equilibrium expectation value,  $\llangle M^2 \rrangle_{\rm eq}$. 
The leading deviation of $\llangle M^2 \rrangle_{\rm eq} $ from $\bsigma^2$ at finite volume comes from fluctuations of long wavelength Goldstone  modes and can be analyzed with the pion  EFT~\cite{Hasenfratz:1989pk}
\st
\label{eq:Hasenfratz}
\llangle M_a M_a\rrangle_{\rm eq}  =  \bsigma^2 \left[1 +  \frac{0.677355 }{f^2 L }  + \mathcal{O}\left((f^2L)^{-2}\right) \right] \, . 
\stp
Here the displayed numerical coefficient and the next term in the series expansion  are known analytically in terms of shape coefficients,  but are of no interest here.
The corrections are organized in inverse powers of $f^2L \sim  L/\xi$, showing again that  $f^2 \propto \xi^{-1}$ in three dimensions. Using \cref{eq:Hasenfratz} we previously extracted the value of $\bsigma$ and $f^2$ as a function of reduced temperature, including subleading corrections~\cite{Florio:2023kmy}, see \cref{tab:subleading}.  

Based on this discussion we will define the correlation length below $T_c$ as 
\st
\xi(\tred)  \equiv \frac{\xi^-}{f^2(\tred)}  \qquad (\tred < 0) \, , \label{eq:corr_belowTc}
\stp
\DT{where $\xi^-=0.479(7)$ is chosen so that the correlation length $\xi(\tred)=\xi^+|\tred|^{-\nu}$ is symmetric around $\tred=0$, up to subleading temperature corrections shown in \cref{tab:subleading}.}
Furthermore, from \cref{eq:Hasenfratz} we get a practical fit to our numerical results used in the plots
\st
\sqrt{\llangle M_a M_a\rrangle_{\rm eq}}  \approx \bsigma \left(1 + 0.706 \frac{\xi }{L }  \right).
\stp

The linearized effective hydrodynamic equations of motion for the pions and axial charge follow from the free energy in \cref{eq:varphifree}, the conservation laws and the Josephson constraint, and a dissipative derivative expansion.  They take the form~\cite{Son:2001ff,Son:2002ci,Grossi:2020ezz}
\begin{subequations}
\label{eq:hydroeqns}
\begin{align}
\partial_t \vec{\varphi}  - \frac{\vec{n}_A}{\chizero} &= \frac{\Gamma }{f^2} \,  \nabla \cdot (f^2  \nabla \vec{\varphi})  \, ,  \label{eq:varphieqns} \\
\partial_t  \vec{n}_A -  \nabla \cdot (f^2 \nabla \vec{\varphi})  &= D \nabla^2 \vec{n}_A  \, .
\end{align}
\end{subequations}
\DT{$\Gamma$ and $D$ are defined by these equations, i.e. they are the dissipative
    parameters of the pion hydro EFT presented in \cref{eq:hydroeqns}.  They scale  with temperature as
$\xi^{2- d/2} \sim \xi^2/\tau_R$ close to the critical point and  they are distinct (and of parametrically different size) from the temperature independent bare parameters,  $\Gamma_0$ and
$D_0$ in \cref{eq:eom}.  However, in mean field theory with $d=4$, $\Gamma$ and $D$ are independent of $\xi$  and  then reduce to $\Gamma_0$ and $D_0$, justifying  the notation.}

The linearized hydrodynamic system in \cref{eq:hydroeqns} is easily solved and the eigen-waves have dispersion relations
\st \label{eq:disp}
   \omega(k) = \pm v k  -  \frac{i}{2} D_A k^2  \, ,
\stp
where $D_A = \Gamma + D$. By systematically analyzing the unequal time correlators  $G_{\pi\pi}(\Delta t , k)$  we previously extracted the  pion dispersion curve numerically with high precision~\cite{Florio:2023kmy}. 

Finally, it is worth recording the ideal $SU(2)_L \times SU(2)_R$ superfluid equations of motion in the fully non-linear regime\footnote{The equations and steps are equivalent to \cite{Son:1999pa}. However, the final equations look rather different because we are using a notation based on $O(4)$ symmetry rather than the $SU_L(2) \times SU_R(2)$ notation adopted by Son.}, following Son~\cite{Son:1999pa}. These equations can be determined by writing down the Poisson brackets between the spins $s_a$ and the conserved charges $n_{ab}$,  which are fixed because the charges generate $O(4)$ rotations~\cite{Rajagopal:1992qz}, e.g. 
 \st
 \left\{n_{ab}({\bm x}), s_c({\bm y}) \right\} = \left(\delta_{ac} \delta_{bd} -  \delta_{bc} \delta_{ad} \right) s_{d}(\x) \, \delta^3(\x - {\bm y}) \, . 
 \stp
 Using the effective spin Hamiltonian in \labelcref{eq:spinfree} and these Poisson brackets,  the equations stemming from $\dot s_a=\{s_s, \mathcal H\}$ and $\dot n_{ab} =\{n_{ab}, \mathcal H\}$  
read
\begin{subequations}
    \label{eq:idealsuper}
\begin{align}
    \partial_t s_a + \frac{n_{ab}}{\chizero} s_b =& 0 \,  ,   \\
    \partial_t n_{ab} +   \partial_i ( f^2 \partial^i s_{[a} s_{b]}) =& 0   \, .
\end{align}
\end{subequations}
In the linearized limit with $s_a \simeq (1, \vec{\varphi})$ the equations reproduce \labelcref{eq:hydroeqns} without the dissipative terms. (The dissipative equations in the non-linear regime are given in \cite{Grossi:2020ezz,Jain:2016rlz}.)
As we will see, the non-linear ideal superfluid theory is responsible for the growth of the chiral condensate following a quench. 

\DT{
\subsubsection{Quasi-particle description of pion waves}
\label{sec:quasi}

In the next section we will study how the medium evolves after a
sudden quench. The pion dynamics post-quench will be analyzed through the time dependence of the equal-time correlation function
\st
G_{\pi\pi}(t, k) \equiv  \frac{V}{3} \llangle \pi^\ell(t, \k) \pi^\ell(t, -\k) \rrangle.
\stp
Now the brackets $\llangle\ldots \rrangle$ notate an average over a rotationally invariant ensemble of initial conditions, determined by the quench. 
Here we will relate the $G_{\pi\pi}(t,k)$ to the pion yield $n_{\pi}(t,k)$, which evolves to equilibrium,  $n_\pi^{\rm eq}(k)$,  after the quench.  

After a global condensate has approximately formed,  the linearized fluctuations of $\phi_a(t, \x)$ are parametrized by the angles $\vec{\varphi}(t, \x)$, and the hydrodynamic equations of motion are given by \cref{eq:hydroeqns}. 
Neglecting the dissipative dynamics,  the linearized equations of motion in \cref{eq:hydroeqns} are similar to a $U(1)$  superfluid --  \cref{eq:varphieqns} is the Josephson constraint relating the time derivative of the phase to the chemical potential, 
$\partial_t \vec{\varphi} = \vec{\mu}_A \equiv \vec{n}_A/\chi_I$~\cite{Son:2001ff}. 

An action reproducing the (ideal) equations of motion given in \labelcref{eq:varphifree} is \cite{Jensen:2012jh,Grossi:2020ezz}
\st
\label{eq:superfluidaction2}
S = \int d^4x \,  \frac{1}{2} \chi_I \, (\partial_t \varphi)^2  - \frac{1}{2} f^2 (\nabla \varphi)^2 \, ,  
\stp
which clearly describes a wave with with dispersion curve $\omega(k)=vk$ and squared velocity, $v^2 = f^2/\chi_I$. 
The action is appropriately normalized:
 the Lagrange density is identified as the superfluid contribution to the pressure~\cite{Jensen:2012jh}. Further,  the canonical momentum conjugate to  the phase  $\vec{\varphi}(t,\x)$ is the charge density $\vec{n}_A(t, \x)$   and  a Legendre transform of the Lagrangian yields the axial part of the free energy,   ${\mathcal H}$,   previously introduced in  \labelcref{eq:varphifree}.    Examining the action and noting that $\vec{\varphi} =\vec{\pi}/\bsigma$,  we see that the rescaled pion field $(\sqrt{\chi_I}/\bsigma) \, \vec{\pi}$ is  canonically normalized,   and may be  expanded in terms of plane wave solutions: 
\st
(\sqrt{\chi_I}/\bsigma) \, \pi^\ell (t, \x) = \sum_{\k } 
a_\k^\ell  \frac{ e^{-i\omega(k) t + i \k \cdot\x }}{\sqrt{2\omega(k) V}}   +  
a_\k^{\ell*}  \frac{ e^{i\omega(k) t - i \k \cdot\x }}{\sqrt{2\omega(k) V}}  \, .  
\stp
Substituting the mode expansion into \labelcref{eq:Gpipi} yields a direct relation between $G_{\pi\pi}$ and the pion yield $n_{\pi}(t, k)$
\st
\label{eq:yieldrelation}
\frac{\chi_I}{\bsigma^2} G_{\pi\pi}(t, k) =  \frac{n_\pi(t, k) }{\omega(k) } \, , 
\stp
where we identified the (isospin averaged) pion yield as
\st
n_{\pi}(t, k) = \frac{1}{3} \sum_{\ell=1}^{3} \llangle |a_\k^{\ell}|^2 \rrangle \, . 
\stp

In writing \cref{eq:yieldrelation} we dropped rapidly oscillating
terms,  $\sim e^{-i2\omega(k) t}$,  in a quasi-particle (or kinetic)
approximation.  If $G_{\pi\pi}(t,k)$ is averaged over a narrow momentum
range $\Delta k$ with $\Delta k \ll k$ and  $v\, \Delta k \, t \gtrsim 1$ these
oscillating terms average to zero.   A more sophisticated
treatment would match the time evolution of the two point
functions of the classical fields onto a Boltzmann
description~\cite{Grossi:2020ezz}, but this analysis is left for
future work. 

When $G_{\pi\pi}$  takes the equilibrium form given in \labelcref{eq:Gpipieq}, $\chi_I G_{\pi\pi}^{\rm eq}/\bsigma^2$ evaluates to 
\st
\label{eq:Gpipieqscaled}
\frac{\chi_I}{\bsigma^2} G_{\pi\pi}^{\rm eq}(t, k)  = \frac{T}{\omega(k)^2} = \frac{n_{\pi}^{\rm eq}(k)}{\omega(k)} \, . 
\stp
In the last step we recognized the classical part of the Bose-Einstein distribution, 
\st
 n_{\pi}^{\rm eq}(k) =  \frac{1}{e^{\omega(k)/T} - 1 } \simeq \frac{T}{\omega(k) } \,  , 
\stp 
and  thus \labelcref{eq:Gpipieqscaled} further corroborates the identification in \labelcref{eq:yieldrelation}.

In what follows  we will monitor the ratio $G_{\pi\pi}(t, k)/G_{\pi\pi}^{\rm eq}(k) = n_\pi(t, k)/n_{\pi}^{\rm eq}(k)$ to evaluate the time-evolution of the soft-pion yield out of equilibrium. 
}
 \section{Out-of-equilibrium dynamics: quenches}
 \label{sec:quenches}

\subsection{Qualitative picture and scaling analysis}\label{sec:scaling}

 In this work we are concerned with the out-of-equilibrium dynamics of Model G,
 particularly the approach to equilibrium of the Goldstone modes. We will start
 by studying the case of instantaneous quenches from the high-temperature restored phase
 to the low temperature broken phase.  By suddenly driving the system 
 with a quench to the broken phase,  we isolate the dynamics that lead to the development of the condensate and its impact on the
 conserved charges and the Goldstone modes.   

\begin{figure}
 \includegraphics[width=0.7\linewidth]{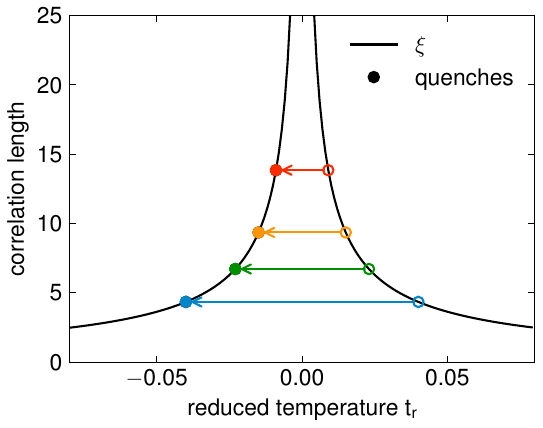}
 \caption{Cartoon of the quench protocol in terms of correlation length $\xi$. The quenches studied in \cref{sec:quenches} transition from the high temperature phase to the low temperature phase,  $\tredini \rightarrow -\tredini$ with $\xi(\tredini)=\xi(-\tredini)$ up to subleading corrections, see \cref{tab:subleading}. \DT{The colors from top to bottom correspond to  
     $\tredini =0.009, 0.015, 0.023, 0.04$ and are correlated with the lines in the upper panels of \cref{fig:VeVquench,fig:quench,fig:sgquenchk}. In the the lower panels of \cref{fig:VeVquench,fig:quench,fig:sgquenchk} and \cref{fig:quench_fixed_k} we vary the lattice size while $\tredini = 0.04$.}
     \label{fig:sim_recap}
 }
 \end{figure}

 Specifically, we initialize our fields in thermal equilibrium specified  by reduced
 temperature $\tred = \tredini>0$, defined in 
 \cref{eq:trOne}.  Then,  at time zero,  the reduced temperature
 is instantaneously changed to $-\tredini$ and the system relaxes to the new equilibrium. The cartoon in \cref{fig:sim_recap} shows the transition  and the equilibrium correlation length above  \labelcref{eq:xidef} and below \labelcref{eq:corr_belowTc} the critical temperature~\cite{Florio:2023kmy}. 

The dynamics of the condensate and its excitations can be analyzed through the time 
evolution of  equal time correlation functions  such as
\st
\label{eq:G0def}
G_0(t) \equiv \frac{V}{4}\sum_{a=1}^4 \llangle M_a(t) M_a(t)  \rrangle \, , 
\stp
which is the non-equilibrium analog of $\chi$. 
The average here and below $\llangle \ldots \rrangle$ reflects an average over the ensemble of initial conditions,  and is distinct from equilibrium average $\llangle \ldots \rrangle_{\rm eq}$ of the previous sections. All other equal time correlation functions such as $G_{\pi\pi}(k)$ and $G_{\sigma\sigma}(k)$ defined in \cref{eq:Gequaltime} become functions of time, e.g., 
\st
G_{\pi\pi}(k)  \rightarrow  G_{\pi\pi}(t, k)  \, . 
\stp
We will focus on the time evolution of equal time correlation functions in this work. 

With our conventions,  correlation functions such as $G_0$ and $G_{\pi\pi}$ are independent of volume when the field is random\footnote{
    More explicitly, our Fourier transform $\phi(\k)\equiv  \tfrac{1}{V} \sum_x e^{-i\k\cdot \x} \phi(\x)$ was defined with a prefactor of
$V^{-1}$. Thus,  $\phi(\k)$ scales as a  volume average of $\phi(\x)$ for $kL\sim 1$. If $\phi(\x)$ is a random field with domains of volume $\ell^d \ll V$, then $\phi(\k)$ is an average  formed with $N_{\rm samp} \propto V/\ell^d$ 
independent samples. Statistically,   this average decreases like $1/\sqrt{N_{\rm samp}}$ or $\sqrt{\ell^d/V}$.
 When defining correlation functions, e.g.,   $G \equiv V\llangle \phi(\k) \phi(-\k)\rrangle $, we multiply by $V$ to have a volume-independent correlation function for random fields.
}, i.e.,  the field is correlated over domains of volume  $\ell^d$ 
 much smaller than $V$. Correlation functions of ``condensate like'' fields,  i.e., fields with domains of volume $\sim V$,  scale linearly in the volume.

\begin{figure}
     \centering
     \includegraphics[width=0.9\textwidth]{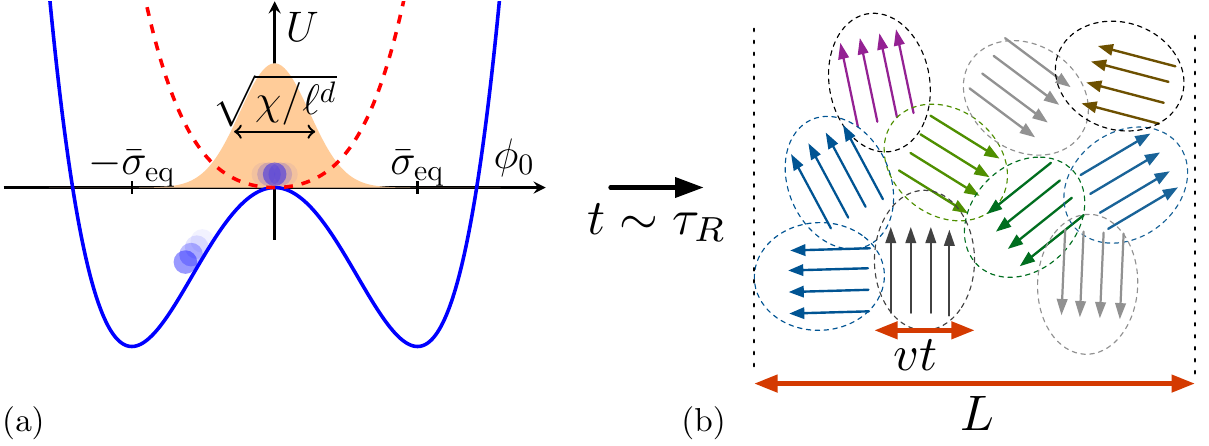}%
     \caption{
(a) Schematic of a quench from the high-temperature phase to the broken phase in a spatial region of size $\ell$,  which is several correlation lengths long (e.g., $\ell \sim 3\xi$). The red dashed line shows the initial potential at temperature $\tredini$, and the blue line shows the post-quench potential. The ball represents a local order parameter \( \bar{\phi}_a(t) \), averaged over the region. The orange Gaussian depicts
the initial thermal distribution for $\bar\phi_a$ before the quench,  with variance \( \chi/\ell^d \). Over a time of order $t\sim \tau_R$, \( \bar{\phi}_a \) evolves 
to the well’s bottom 
in a random direction, forming a local equilibrated domain of size $\ell$. 
(b) Field configuration for \( t \gg \tau_R \) in a region of size \( L \gg \ell \sim \xi \), showing randomly oriented domains of the chiral condensate.  The domains  merge over a time of order \( t \sim L/v \) with dynamics governed by the ideal superfluid equations given in \labelcref{eq:idealsuper}.
}
     \label{fig:drawing1}
 \end{figure}

 The quench dynamics is illustrated schematically in \cref{fig:drawing1}.
 Consider a local spatial region of length $\ell$, several correlation lengths long.  In the restored phase, before the quench,    the potential is shown by the red dashed line, while after the quench the potential takes the familiar wine-bottle form.
 The order parameter  averaged this region is notated $\bar{\phi}_a$, and its value before the quench fluctuates around zero,  with variance order ${\chi}/\ell^d$ (see figure).  
After the quench,    $\bar\phi_a$ now sits at the
 top of the hill and is unstable. Over a time of order $t\sim \tau_R$,  fluctuations  source an unstable
 growth of the local order parameter, which rolls down the hill in a random direction, forming a locally-equilibrated domain.
 At this stage, the global condensate is still approximately zero,  since the local condensate orientations are random, as shown in \cref{fig:drawing1}(b).  Over a much longer timescale these the domains begin to merge to form the global condensate. As discussed further below, since the distances and times involved are much larger than the correlation length $\xi$ and the relaxation time $\tau_R$ respectively, hydrodynamics is the right tool to describe the growth in this stage. The appropriate  theory was given in \cref{sec:hydroGM} and describes a non-abelian superfluid with broken $SU(2)_L \times SU(2)_R$ symmetry.

Scaling considerations let us go beyond this qualitative understanding and make sharp predictions on the effect of this non-equilibrium evolution on the correlators.
At time $t=0$ (the ``top of the hill''), the system is in equilibrium in the restored phase and the correlator reflects the  magnetic susceptibility:
 \begin{equation}
     \left. G_0(t) \right|_{t=0^-} = \chi \sim \xi^{2-\eta} . \label{eq:Gpipik0I}
 \end{equation}
This means that at $t=0$ the fluctuations of $M_a$ are of order $\sim \sqrt{\chi/V}$.  By contrast, at late times
a single randomly oriented domain is formed with $M_a = \bsigma \nhat_a$. This implies that
  \begin{equation}
      \lim_{t\rightarrow \infty} G_{0}(t) = \frac{V}{4}\bsigma^2 \sim  \xi^{2-\eta} \left(\frac{L}{\xi}\right )^d. \label{eq:Gpipik0III}
 \end{equation}
 Thus,  $G_0$ grows by a factor which is a ratio of these limiting values:
 \st
  \mbox{amplification} = \frac{ V\bsigma^2}{4\chi}   \sim \left(\frac{L}{\xi}\right)^d \, . 
 \stp
 This factor will appear frequently below. 

 The scaling hypothesis quite generally dictates the dependence of $G_0$ on time and the correlation length
 \begin{align}
     G_0(t, \xi, L) =  \xi^{2-\eta} {\mathcal F}(t /\tau_R, \xi/L) , \label{eq:scaling_hypothesis}
 \end{align}
 where here and below ${\mathcal F}$ denotes some dimensionless function, distinguished by context and arguments.   For $\xi/L \ll 1$,  \cref{eq:scaling_hypothesis} suggests a  naive scaling form 
 \begin{align}
     \label{eq:G0early}
     G_0(t, \xi) =  \xi^{2-\eta} {\mathcal F}(t /\tau_R)      \qquad   \mbox{for $t\sim \tau_R$} \, . 
 \end{align}
 The naive scaling form in \labelcref{eq:G0early}, which is valid for $t\sim \tau_R$ and is independent of the system size,  describes the formation of the local domains as discussed in \cref{fig:drawing1}(a). However,  this naive scaling drops secular terms which become important at late times when $t \sim \tau_R (L/\xi)$.

 Indeed, as described above  $G_0$ must be of order $(L/\xi)^d$ at late times. We anticipate that the condensate will grow in the time it takes for a pion to propagate across the box,  $t \sim L/v \sim \tau_R (L/\xi)$. Recognizing these scalings, we reshuffle  the parameters in \cref{eq:scaling_hypothesis} into a new dimensionless function 
\st
G_0(t, \xi, L) = \xi^{2-\eta} \left(\frac{L}{\xi}\right)^d {\mathcal F}(vt/L,  \xi/L) \, .  \label{eq:scaling_hypothesis1}
\stp
So far, this is a trivial rewrite of \cref{eq:scaling_hypothesis}. 
Now, however, the new function is expected to be regular and order one as $L\rightarrow \infty$ with $vt/L\sim 1$ fixed,   and its dependence on $\xi/L$ can be dropped. 
Thus, at large volumes and at late times $v t/L \sim 1$ we expect a scaling form for the time evolution of the magnetization
\st
G_0(t, \xi, L) = \xi^{2-\eta} \left(\frac{L}{\xi}\right)^d {\mathcal F}(v t/L) \,,  \qquad \mbox{ for $t\sim L/v$.} \label{eq:scaling_hypothesis2}
\stp

The scaling form in \cref{eq:scaling_hypothesis2} is  valid at intermediate times  $t\sim L/v$  when ballistic transport of pions is operative as indicated in \cref{fig:drawing1}(b). 
However, the scaling form does not capture the late time equilibration dynamics of the Goldstone modes, which happens on the diffusive time scale,  $t\sim \tau_R(L/\xi)^2 \sim  (L/v) (L/\xi)$.  
In \cref{eq:scaling_hypothesis2} the approach to equilibrium  at late times has been discarded  by ignoring possible secular terms in passing from \labelcref{eq:scaling_hypothesis1} to \labelcref{eq:scaling_hypothesis2}. 

At early times, but still with $t\gg \tau_R$,  the global condensate has not formed. Then, $G_0$ is  a correlation function of a random field, a field    consisting  of domains of volume $\sim (v t)^d \ll V$, i.e.  \cref{fig:drawing1}(b) with small domains. In this regime $G_0$ will be independent of the system volume.  As a result,  for $vt/L\ll1$  we must have  $\mathcal F(vt/L) \propto (vt/L)^d$  to cancel the leading power of $L^d$ in \labelcref{eq:scaling_hypothesis2}. Using the estimate $v \sim \xi/\tau_R$,  we thus extract the early time behavior of $G_0$ 
in the overlap region of \labelcref{eq:G0early} and \labelcref{eq:scaling_hypothesis2}
\st
\label{eq:overlap}
G_0(t, \xi, L) \propto \xi^{2-\eta} \left(\frac{t}{\tau_R}\right)^d     \qquad  \tau_R \ll t \ll  \frac{L}{v}  . 
\stp

The timescale $L/v$ of the scaling regime in \labelcref{eq:scaling_hypothesis2} is  much longer than the relaxation time $\tau_R$.  The distance scale $vt\sim L$ is much longer than the correlation length $\xi$. Thus, the scaling regime in \labelcref{eq:scaling_hypothesis2} is a hydrodynamic regime.  The appropriate hydrodynamic theory is given by an $SU(2)_L\times SU(2)_R$ superfluid described in \cref{sec:hydroGM}, which accounts for the broken symmetry. As discussed above,  the scaling form in \labelcref{eq:scaling_hypothesis2}  ignores the dissipative dynamics and thus,  the relevant hydrodynamics is the non-dissipative (but non-linear) superfluid dynamics given in \labelcref{eq:idealsuper}. In the future, it should be possible to simulate these equations directly to describe the condensate growth.

Scaling forms for condensate growth, similar to \cref{eq:scaling_hypothesis2}, are common in analyses of coarsening in other critical systems~\cite{tauber2014critical,onuki2002phase}. However, the linear growth of the domain size, $\sim vt$, is a distinctive feature of the superfluid limit of Model G, ultimately arising from its reversible Poisson bracket structure.

The critical scaling in \cref{eq:scaling_hypothesis2}  can also be extended to small but finite momenta.  
For simplicity, assume that the \( \k = {\bm 0} \) mode, \( 
\phi_a(\mathbf{0}) \equiv M_a \), is pointing in the \( \nhat_a = (1,0,0,0) \) direction. We
then analyze order parameter fluctuations \( \phi_a(\k) =
(\sigma(\k), \vec{\pi}(\k) ) \) at momenta satisfying \( kL
\sim 1 \).  For $t \lesssim L/v$, at the boundary of applicability of \labelcref{eq:overlap},  pion waves have not 
fully traversed the box or other distances of order \( \sim 1/k \), see \cref{fig:drawing1}(b). As a result, the zero mode and other long wavelength modes have  not had  time to dramatically influence the  each other and form a global condensate. Both are primarily produced from random superpositions of smaller scales (see \cref{fig:drawing1}(b)).  Consequently, the fluctuations of  $\phi_a(\k)$ 
will be the same order of
magnitude as the zero mode, $\phi_a({\bm 0})$. In addition,  the decomposition of $\phi_a(\k)$ into $(\sigma(\k), \vec{\pi}(\k))$, which is based on the direction of the zero mode $\nhat_a$,  is not physically significant and all components of $\phi_a(\k)$ are the same order of magnitude.
Thus, the pion and sigma correlation functions will follow a scaling form similar to the zero mode, for example: 
\begin{equation}
\label{eq:scaling_hypothesis3}
G_{\pi\pi}(t, \xi , k , L)  = \xi^{2-\eta} \left(\frac{L}{\xi} \right)^d \mathcal F(v k t , kL) .
\end{equation}
Here we recognized that $\mathcal F(vkt, kL)$  is an order-one  function when $kL\sim 1$ and $vkt \sim 1$. For later convenience,  we replaced $1/L$ in the time argument of \labelcref{eq:scaling_hypothesis2} with the wavenumber $k$ of the relevant mode,  writing $vkt = (vt/L)\, (kL)$. 

Now in
the intermediate range
\begin{equation}
   L^{-1} \ll  k \ll  \xi^{-1} \, , 
\end{equation}
and for times $t \sim (vk)^{-1} $, which are short compared to $L/v$,  the correlation function must again become independent of the system volume. Thus,  we anticipate that for $kL\gg 1$, the function $\mathcal F(vkt, kL)$ simplifies,   $F(vkt, kL)\rightarrow  \mathcal F(vkt)/(k L)^d$,  leading to the scaling form 
\begin{equation}
    G_{\pi\pi}(t, \xi, k) = \frac{\xi^{2-\eta}}{(k \xi)^d } \mathcal F( v kt ) . \label{eq:Gpipik}   
\end{equation}
Again, this description is only valid for $t\sim (vk)^{-1}$ and does not capture the diffusive behavior at late times,  $t\sim \tau_R /(k\xi)^2  \sim 1/(vk)\, 1/(k\xi)$.

The prediction in \cref{eq:Gpipik}, which  follows from scaling and the dynamics of Model G,  is quite striking. After noting that the equilibrium correlation function in \cref{eq:Gpipieq} scales as $G_{\pi\pi}^{\mathrm{eq}}\sim  \xi^{2-\eta} \left({k\xi}\right )^{-2}$,  we see that the equation  predicts  a significant  enhancement over  equilibrium. \DT{In the quasi-particle regime  discussed in \cref{sec:quasi} where $vkt\gtrsim 1$, this translates to the parametric enhancement of soft pion spectra}
\begin{align}
    \frac{G_{\pi\pi}(t,k)}{G_{\pi\pi}^{\mathrm{eq}}(k)} \simeq \frac{n_{\pi}(t,k)}{n_{\pi}^{\mathrm{eq}}(k)}\sim \frac{1}{k\xi}  \qquad  \mbox{with $k\xi\ll 1$}.
\end{align}
This enhancement remains substantial over an extended timescale, \DT{ $\tau_R/(k\xi) \lesssim  t \ll \tau_R /(k \xi)^2$}, demonstrating a parametrically long-lived deviation from equilibrium.  

  \begin{figure}
     \centering
     \includegraphics[width=0.7\linewidth]{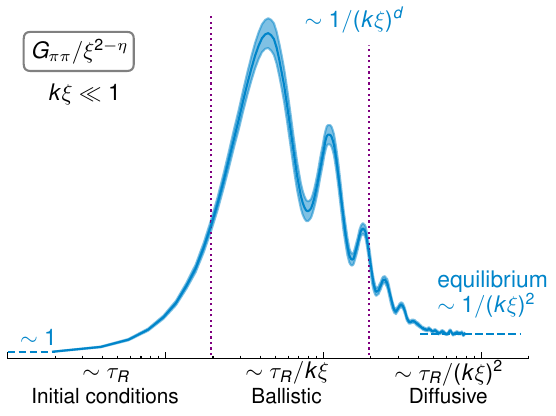}
     \caption{A sketch of different stages in the typical evolution of the (pion) two-point function: early evolution setting the initial conditions $t\sim \tau_R$, rapid growth until the maximum, and slow decay to its equilibrium value in an oscillatory manner. See \cref{sec:scaling} for detailed discussion.}
     \label{fig:Gpipisingle}
 \end{figure}
 
All of the scaling predictions discussed here are observed in our numerics, as presented in the next section. Before moving on, we summarize the situation with \cref{fig:Gpipisingle}, which shows
 a schematic of a pion correlator $G_{\pi\pi}$ obtained from our numerics on a logarithmic time axis. The three relevant time scales $t\sim\tau_R$ (initial conditions), $t\sim \tau_R/(k\xi)$ (ballistic transport) and $t\sim \tau_R/(k\xi)^2$ can be clearly seen from the behavior of the correlator. The $1/k\xi$ enhancement over  equilibrium  during the ballistic phase is particularly striking. Again, since the timescale in the ballistic regime is long compared to $\tau_R$,  the  superfluid hydrodynamic equations reflecting the broken $O(4)$ symmetry in \labelcref{eq:idealsuper}   determine the dynamics of this regime.

 \subsection{Lattice simulations of quenches}
 \label{sec:lattice_quench}

 In this section, we gather all our different numerical
 simulations of quenches and show that the above scaling
 analysis holds.\DT{ To give a sense of the numerical effort,
 consider the orange quench in \cref{fig:sim_recap}, 
which  produces orange curves in the top two panels of \cref{fig:VeVquench}.
The relaxation time,   $\tau_R \propto \xi^{d/2}$,  is of order $\tau_R \sim 150\,  (a^2/\Gamma_0)$.  The box size and correlation length are of order $(L/\xi)\sim 20$ and $\xi \sim 10 \, a$  respectively. The relaxation time is large compared to $a^2/\Gamma_0$  because we are close to the critical point and the dynamics shows critical slowing down.  A typical event is first thermalized by evolving for $t \sim 1200 (a^2/\Gamma_0)$, with timesteps of $0.24\, (a^2/\Gamma_0)$ as discussed above. After the quench,  the system is evolved for $\sim 50 \, \tau_R$,  which  is ample time for the longest wavelength Goldstone mode to relax to equilibrium for $L/\xi \sim 20$.  The relaxation time for the longest Goldstones is of order $\sim \tau_R L^2/(2\pi \xi)^2$. The events are run on single NERSC CPU node consisting of 128 cores.  The curves in \cref{fig:VeVquench} represent an average over $\sim 40$ events. }

 \subsubsection{Magnetization evolution}

 \begin{figure}
\includegraphics[width=\textwidth]{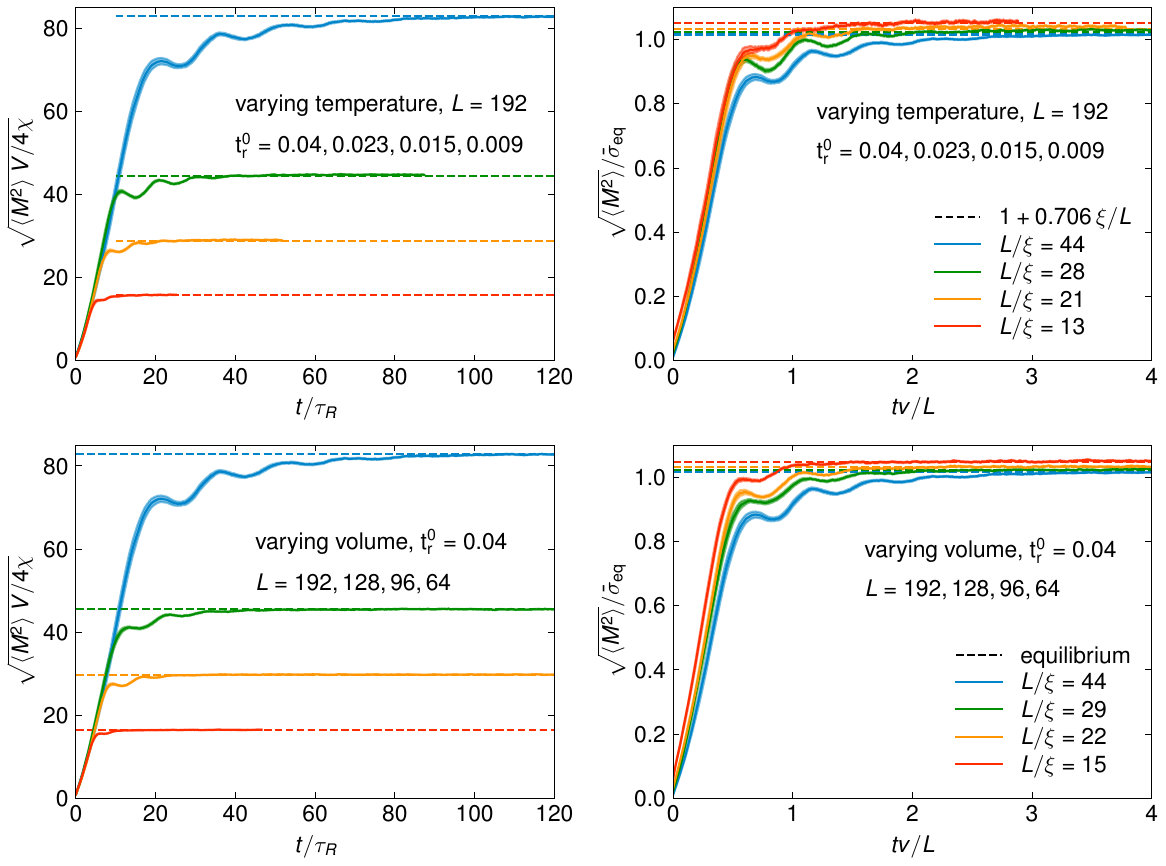}
 \caption{Time evolution of magnetization for symmetric quenches in \cref{fig:sim_recap}, with (top) different
 starting $\tredini$, but same lattice size $L=192$ and (bottom) with $\tredini=0.04$ and different lattice
 sizes. All curves are normalized by the initial equilibrium value, \cref{eq:chi}. Colors indicate the system size in units of correlation length $L/\xi$. The time is given in units of relaxation time $t/\tau_R\propto t/\xi^{d/2}$ and pion ballistic propagation time $L/v$. Dashed lines show the expected equilibrium value with the leading order
 finite-size correction \cref{eq:Hasenfratz}.\label{fig:VeVquench}}
 \end{figure}

First, in \cref{fig:VeVquench} we show the time evolution of the
magnetization, $\sqrt{\langle M^2(t)\rangle}$. Here and below,
the brackets $\llangle \ldots \rrangle$ denote an average over
events. Before the quench, when the system is in equilibrium,
$\sqrt{\langle M^2\rangle} = \sqrt{4\chi/V}$, where $\chi(\tred)$
is the magnetic susceptibility above $T_c$. Accordingly, in
\cref{fig:VeVquench} (left column), the $y$-axis is normalized by
this equilibrium value so that all curves start at the same
point. After the quench, the system relaxes and approaches its
new equilibrium value given by $\bsigma(\tred)$  up to finite corrections, \cref{eq:Hasenfratz}. In
\cref{fig:VeVquench} (right column), the $y$-axis is normalized
by $\bsigma$. \DT{ Here and throughout, the rescalings use the parametrizations displayed
in \cref{tab:subleading} for $\chi(\tred)$, $\bsigma(\tred)$, and other quantities. }

 On the top panels of \cref{fig:VeVquench} we keep lattice size $L=192$ fixed and vary the initial reduced temperature $\tredini$ with line colors correlated with \cref{fig:sim_recap}. On the lower panels, we fix $\tredini=0.04$, but change the lattice size. The almost identical curves illustrate the fact that the only relevant parameter is the system size in units of correlation length, i.e., $L/\xi$, which is the same in both panels. 
 
 The left panels focus on early time dynamics. By rescaling the $x$-axis by $\tau_R$\footnote{\DT{We assume use the relaxation time $\tau_R$ is symmetric around $\tred=0$ and use the parametrization for $T>T_c$ in \cref{tab:subleading}.}}, all curves collapse at early times. This corresponds to the first regime of interest $t\sim \tau_R$ mentioned in the previous sections, \cref{eq:G0early,eq:overlap}. 
The right panels focus on the ballistic regime $t\sim L/v$ captured by the scaling analysis of the previous section. Rescaling the y-axis by $\bsigma$, we expect to see all curves eventually to diffuse back to one, up to the finite volume corrections predicted by the poin EFT and given in \labelcref{eq:Hasenfratz}. More interestingly, rescaling the time axis by $v/L$ confirms the main hypothesis in our scaling analysis: intermediate time scales are described by a universal, regular function $\mathcal F(tv/L)$. This can be seen by the data collapse, which is hindered to fully happen only by the finite volume corrections of order $\xi/L$.\footnote{Note that we could have also rescaled the $y$-axis by $M^2_{\rm eq}$ instead of $\bsigma$ and obtained a better collapse. We refrained from doing so as strictly speaking the scaling analysis presented in the previous section does not control the finite volume corrections responsible for the difference between these two quantities.}

 \subsubsection{Pion and sigma correlation functions}

 \begin{figure}
 \includegraphics[width=\columnwidth]{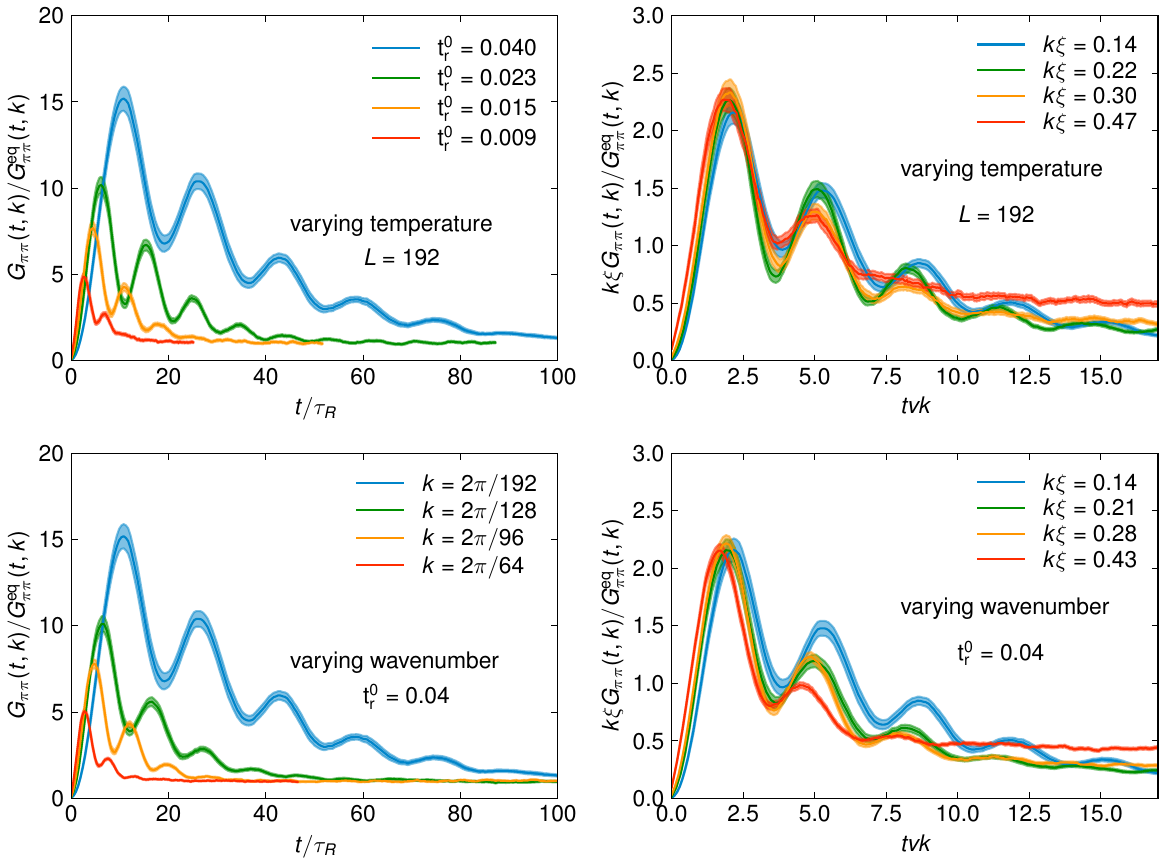}
 \caption{(top left) Enhancement of equal time pion-pion field correlator over thermal
 equilibrium expectation during symmetric quench for different starting $\tredini$
 and wavenumbers $k=2\pi/L$, where  $L=192$.  The time is given in units of relaxation time $t/\tau_R\propto t/\xi^{d/2}$. (top right) The same plot multiplied
 with $k\xi$ on $y$-axis and  time given in units of pion ballistic propagation time $L/v\sim \tau_R (L/\xi)$, where $v$ is the pion velocity at corresponding
 $\tred=\tredini$. (bottom left and right) Enhancement of pion-pion field correlator over thermal
 equilibrium expectation during symmetric quench from $\tred=0.04$ to
 $\tred=-0.04$ for different wavenumbers $k= 2\pi/L$, where
 $L=192,128,96,64$.\label{fig:quench}}
 \end{figure}

 Next, in \cref{fig:quench} we show the evolution of pion-pion correlation
 function $G_{\pi\pi}(t,k)$ for the lowest momentum mode $k=2\pi/L$, normalized by its equilibrium value. Analogously to \cref{fig:VeVquench}, we vary the reduced temperature (upper row) and lattice size (lower row). Again, only the value of $k\xi$ is relevant to characterize the behavior of the correlator, which is the same in all panels. Normalizing by $\tau_R$, we see again universality in the initial regime shown by the left two panels, although the scaling is not quite as good as the condensate previously shown in \cref{fig:VeVquench}. 

 The ballistic transport phase is further studied in the second column, where we now multiply the correlator ratio by $k\xi$ and plot it as a function $vkt$. (Since the equilibrium correlator scales as $G_{\pi\pi} ^{\rm eq} \propto 1/k^2$, multiplying $G_{\pi\pi}/G_{\pi\pi}^{\rm eq}$ by $k\xi$ removes the leading $1/(k\xi)^d$ factor from the scaling form in \labelcref{eq:Gpipik}.)
 The data collapse is impressive, fully confirming the scaling prediction. The $1/k\xi$  enhancement over the equilibrium value is clear.

 \begin{figure}[tbp]
 \includegraphics[width=\columnwidth]{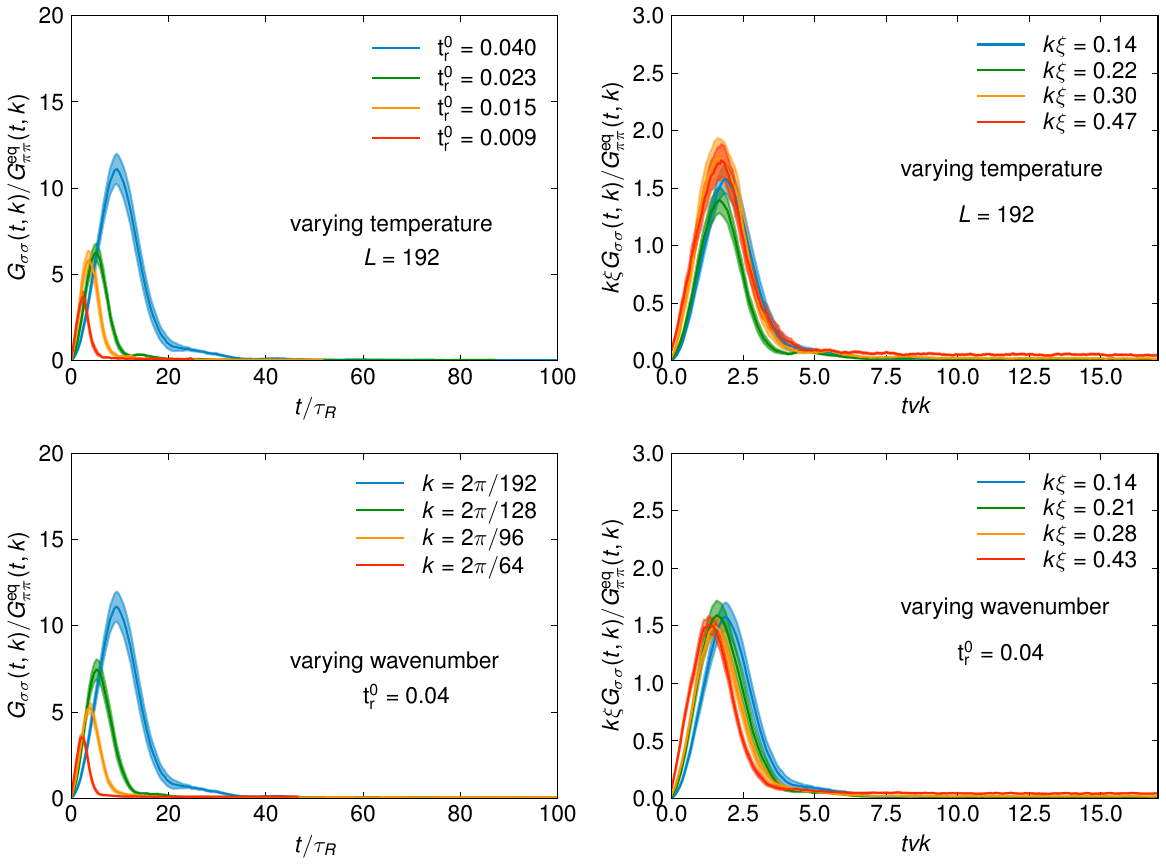}
 \caption{(top left) Enhancement of the equal time sigma-sigma field correlator over thermal
 equilibrium expectation of pion-pion correlator during symmetric quench for different starting $\tredini$
 and wavenumbers $k=2\pi/L$, where  $L=192$.  The time is given in units of relaxation time $t/\tau_R\propto t/\xi^{d/2}$. (top right) The same plot multiplied
 with $k\xi$ on $y$-axis and time given in units of pion ballistic propagation time $L/v \sim \tau_R (L/\xi)$, where $v$ is the pion velocity at the corresponding $\tred=-\tredini$. (bottom left and right) Enhancement of sigma-sigma field correlator over thermal
 equilibrium expectation during symmetric quench from $\tred=0.04$ to
 $\tred=-0.04$ for different wavenumbers $k=2\pi/L$, where
 $L=192,128,96,64$.\label{fig:sgquenchk}}
 \end{figure}

 In \cref{fig:sgquenchk} we repeat the same analysis for the sigma-sigma correlation function, $G_{\sigma\sigma}(t,k)$. In equilibrium, this correlator relaxes to nearly zero. Therefore we normalize the $y$-axis with the \textit{pion} equilibrium correlator $G_{\pi\pi}^\text{eq}=
 k^{-2}\bar{\sigma}^2/f(\tred)$, instead of $G_{\sigma\sigma}^{\rm eq}$. 
 As argued in the scaling section, the initial dynamics proceeds in the same way until the condensate is well formed, further motivating scaling $G_{\pi\pi}$ and $G_{\sigma\sigma}$ by the same factor. The drastic difference between the pion and sigma channel is of course that the sigma mode does not have long lived modes, and thus quickly decays back to equilibrium. Rescaling the time with the ballistic time $v/L$ and the amplitude with $1/(k\xi)$ leads to a reasonable data collapse. 
 
\begin{figure}
    \centering
    \includegraphics[width=\linewidth]{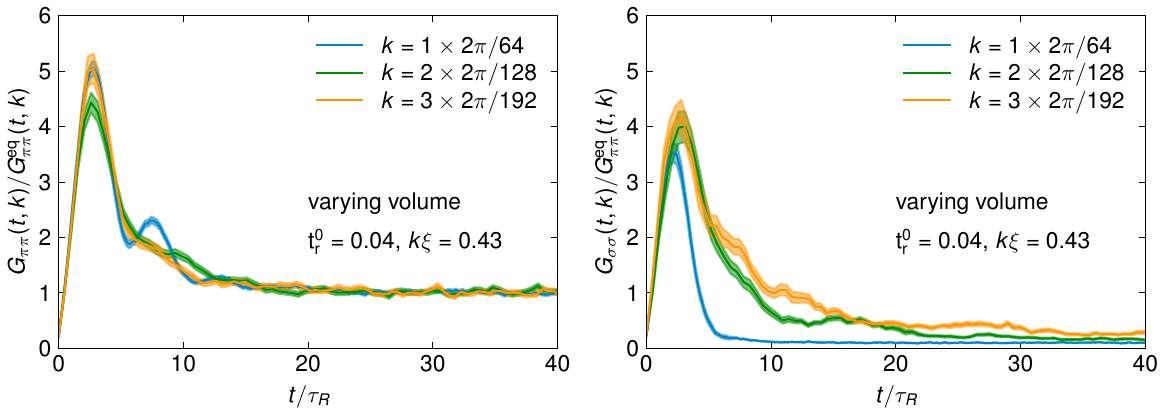}
    \caption{
    (left) Enhancement of the equal time pion-pion field correlator over thermal
 equilibrium expectation during symmetric quench from  $\tred=0.04$ to
 $\tred=-0.04$ for fixed wavenumber $k=\pi/32$, but different lattice sizes  $L=192,128,64$. The time is given in units of relaxation time $t/\tau_R\propto t/\xi^{d/2}$.
 (right) The same plot for sigma-sigma correlator.
    }
    \label{fig:quench_fixed_k}
\end{figure}

Next we study the dependence of the pion and sigma correlations on $kL$ at fixed $k\xi$.  
In \cref{fig:quench_fixed_k} we show pion-pion and sigma-sigma correlation functions for fixed temperature and fixed wavenumber $k=n\times 2\pi/L$, but varying system size. That is for lattice sizes $L=64, 128$, and $192$ we show the $n=1$, $n=2$, and $n=3$ modes correspondingly. 
\cref{fig:quench_fixed_k}(a) shows the pion channel. The resulting curves are approximately independent of $kL$ at fixed $k\xi$. However, close inspection shows that the higher Fourier modes, e.g.,   $k=3(2\pi)/L$,  exhibit weaker oscillations than the $n=1$ curve.  Indeed, at larger $n$,  a specified wavenumber can mix with many other modes of similar frequency,  leading to decoherence. In the limit where $kL\gg 1$ with fixed $k\xi$,  we expect that the oscillations seen in the $n=1$ mode of \cref{fig:quench}  will disappear entirely.  

\cref{fig:quench_fixed_k}(b) shows the sigma channel. In general the $\sigma$ can mix with two pions, making it difficult to define the screening mass $m_{\sigma}$ below  $T_c$~\cite{Engels:2009tv}. In real time,  if the frequencies are commensurate, this mixing introduces slowly decaying contributions to the sigma channel. 
For $n=2$ and $n=3$ modes we see significant corrections compared to $n=1$ behavior. Indeed, the sigma correlator decays on the same time scale as the pion-pion correlation function shown on the left, suggesting that this mixing is at work.

 \subsection{Mean field description of quenches}\label{sec:mean-field-quench}
 \subsubsection{Overview and preliminaries}
 \label{sec:meanintro}
 In this section, we will analyze the quench dynamics in mean field theory, which provides a simple analytic model for the dynamics.  However, because of its simplicity the mean field description fails in several respects, both qualitatively and quantitatively. 

 In mean field theory, the zero mode is separated from the remaining modes, which are then linearized around the zero mode, which is in general time dependent.  The equations of motion are presented in  \cref{app:meanfield} and \cite{Grossi:2021gqi}  where the equilibrium properties of the model have been studied previously. 

 Briefly, to set notation,  the static susceptibility above $T_c$ follows from the mean field free energy, which is the quadratic part of the Landau-Ginzburg free energy in \cref{eq:Landauginzburg}
 \st
 \label{eq:meanfieldsusc}
 G_{\phi\phi}(k) =  \frac{1}{k^2 + m_0^2}  =  \frac{ \chi }{(k\xi)^2 + 1 } 
 \stp
 In the last step  we identified the mean field correlation length
\st
\xi \equiv \frac{1}{|m_0|} \propto \tred^{-1/2} 
\stp
and  the mean-field susceptibility 
\st
\label{eq:meanfieldchi}
\chi  \equiv  \lim_{k \rightarrow 0}  G_{\phi\phi}(k) = \frac{1}{m_0^2} \equiv \xi^{2}
\stp
The condensate below $T_c$ follows by minimizing the potential $U(\phi)$ in \cref{eq:potential}
\st
\bsigma =  \frac{|m_0(\tred)|}{\sqrt{\lambda}}   \propto  \xi^{-1}
\stp
From these relations we deduce familiar mean field exponents
 $\beta =  \nu = \tfrac{1}{2}$  and    $\eta = 0$. 
 These are consistent with hyperscaling in \cref{tab:num_values_fixed} only in $d=4$.  

In mean field the dynamical parameter $\mGamma$  is constant and the typical relaxation time is 
\st
    \tau_R \equiv  \frac{1}{\mGamma |m_0|^2}  \propto \xi^{2}\label{eq:meantauR}
\stp
Again, this is consistent with the scaling $\tau_R \propto \xi^{d/2}$ only in $d=4$. We will measure time in units of $\tau_R$ for our mean field analysis.  The pion dispersion curve in mean field takes  a familiar form of \cref{eq:dispersion} ~\cite{Grossi:2021gqi}. The decay constant in mean field is $f^2 = \bsigma^2 \propto \xi^{-2}$,  which is consistent with the previous section in $d=4$. 
With this scaling,  the velocity in \cref{eq:velocity}  
 scales as $\xi/\tau_R$ as before, with a coefficient that depends on  $\chizero$.    
 As a result, for modes of wavelength $L$ the timescales 
 $\tau_R$,\,  $\tau_R (L/\xi)$ and $\tau_R (L/\xi)^2$ identified in \cref{eq:timescales}
 remain valid.

As discussed previously, the qualitative picture of condensate growth following a quench is discussed in \cref{fig:drawing1}.  Following a quench, the system is unstable at ``the top of hill''   and the condensate subsequently grows to its equilibrium value.   We will first analyze the condensate correlator $G_0(t) = V \llangle M^2 \rrangle /4$,  defined in \cref{eq:G0def}.   As emphasized in \cref{sec:scaling},  $G_0$ needs to grow from $\chi$ to $V\bsigma^2/4$ increasing by a factor
\st
\label{eq:amplification}
\mbox{amplification} \equiv \frac{V\bsigma^2}{4\chi}  = \frac{1}{4\lambda} \left(\frac{L}{\xi}\right)^d  \, . 
\stp
In the last step we assumed $d=4$,  where $\lambda$ is dimensionless,  and the amplification factor scales as $(L/\xi)^d$.  Outside of $d=4$,  \cref{eq:amplification} is simply $L^d/\lambda \xi^4$ and the mean-field results will not exhibit scaling.
As we will  see  in the next section, in mean-field 
the growth of the condensate takes the scaling form in $d=4$
\st
\frac{G_0}{\chi}  = \left(\frac{L}{\xi}\right)^d \mathcal F(t/\tau_R, \xi/L )   \, ,  
\stp
as in the critical dynamics. 

However, there are striking differences between the mean field and critical dynamics.  Specifically,  the condensate growth is exponential, rather than ballistic, which reflects the instability at the ``top of the hill''. 
The e-folding time for the exponential growth is $\tau_R$  and it takes $\sim \ln(V\bsigma^2/\chi)$ e-foldings before $G_0$ can grow from its initial value of $\chi$ to its final value, $V\bsigma^2/4$.  Thus,  since $G_0$ is proportional to $M^2$, we will  define a timescale $t_L$ characterizing the growth of the condensate $M$:
\st
\label{eq:t0def}
t_L \equiv \frac{\tau_R}{2}  \ln\left(\frac{V\bsigma^2}{8\chi} \right)  \sim  \tau_R \ln((L/\xi)^{d/2}) \, ,
\stp
Here the factor of eight in the logarithm (as opposed to four) is without significance and reflects the details of the computation given below. 
The timescale $t_L$,  which depends only logarithmically on $L$,  should be contrasted with the ballistic timescale $\tau_R (L/\xi)$ of the previous section.

 \subsubsection{Growth of magnetization}\label{sec:volume}

 In a mean field description we first look at the time evolution of the zero mode, which we find by substituting a spatially homogeneous field $\phi_a(t, \x) = M_a(t)$ in \cref{eq:eom1} and averaging the equation over  volume
 \st
 \label{eq:basiceom}
 \partial_t M_a = - \Gamma_0 (m_0^2 + \lambda \sigma^2) M_a  + \xi\, ,   \qquad  \llangle \xi_a(t) \xi_b(t') \rrangle = \frac{2 \mGamma }{V}  \delta_{ab} \delta(t - t')  \, . 
 \stp
 Here and below $\sigma^2 = M_a M_a$.  In the quench setup,  this equation 
 should be solved with initial conditions drawn from the equilibrium distribution at temperature  $\tredini$,   before evolving the system at temperature $-\tredini$ for $t> 0$ with $m_0^2 < 0$. 

 The probability distribution for the magnetization $M_a$ in the symmetric phase $T>T_c$ is Gaussian provided $L \gg \xi$:
 \begin{equation}
     P(M_a) = \mathcal{N}\exp\left[-  \frac{V}{2\chi}  M_a
     M_a\right].
 \end{equation}
Concretely,
 \begin{align}
 \langle M_a \rangle &= 0, \\ 
     \langle M_a M_b \rangle &=\frac{\chi}{V}\delta_{ab} , \label{eq:gaussian-var}
 \end{align}
 with the higher-order moments further suppressed by powers of the volume. 
 At early times the field amplitude is small and the non-linear terms can be dropped. The probability distribution of $M_a$ is well characterized by the variance. On the other hand, at early times the noise plays an important role in the evolution. The equation of motion for the two point function which follows from \labelcref{eq:basiceom} and in this limit reads
 \begin{equation}\label{eq:schwinger-dyson}
     \partial_t \langle M_a M_b \rangle = 2 \mGamma |m_0^2| \,  \langle M_a M_b \rangle +
     \frac{2 \mGamma }{V} \delta_{ab},
 \end{equation}
 which is solved by exponential growth:
 \begin{equation}
     \langle M_a M_b \rangle=\delta_{ab}\frac{\chi }{ V} \, (2e^{2\, t/\tau_R}-1).  
 \end{equation}
where we recall that $\tau_R=\chi/\Gamma_0$.
 In terms of the schematics of \cref{fig:drawing1}, this stage corresponds to the initial time when the field starts rolling down the potential.  The squared amplitude evolves at late times as 
 \st
 \label{equation:shortimes}
 \sigma^2 = \langle M_a M_a \rangle = 2 \left(\frac{4\chi}{V} \right) \, e^{2\, t/\tau_R} \, . 
 \stp

 After these initial stages, when $\sigma^2 \gg \chi/V$, the non-linearities cannot be neglected from the mean-field equations. However, the intrinsic effect of the noise is subdominant and suppressed by volume.
 The appropriate simplification of \cref{eq:basiceom} in this limit is
  \st
 \label{eq:dt_meanfield}
  \partial_t  \sigma^2 = -2\Gamma_0  m^2(\sigma)\sigma^2 ,
 \stp
 where $m^2(\sigma) = -|m_0^2| + \lambda
   \sigma^2$.
 The solution is given by
 \begin{align}
  \sigma^2(t)=
 \frac{\bsigma^2}{  e^{ -2 (t - t_L)/\tau_R }+1}\;,
 \label{eq:sigmasol}  
 \end{align}
 where,  at this point,  $e^{2t_L/\tau_R}$ is an integration constant, which remains to be fixed.

The integration constant is chosen so that late time solution  matches the exponential growth of the early time behavior in a region of overlap. Comparison of \labelcref{equation:shortimes} and \labelcref{eq:sigmasol} shows that 
  \begin{equation}
       \frac{V\bsigma^2}{8\chi}  = e^{2\, t_L/\tau_R}\; . 
\end{equation}
as anticipated in the text surrounding \cref{eq:t0def}. 
The correlation function $G_0/\chi$ takes the form 
\st
\frac{G_0(t, \xi , L ) }{\chi}   = \frac{1}{4\lambda} \left(\frac{L}{\xi}\right)^d   \;  \frac{e^{2(t - t_L)/\tau_R }}{1+e^{2(t - t_L)/\tau_R } } \, . \label{eq:meanfieldG0} 
\stp

To summarize,   we have shown that in mean field the global condensate is formed on a timescale of $t_L \sim \tau_R \ln((L/\xi)^2)$.  This timescale should be contrasted with the scaling results and our numerical simulations presented in \cref{fig:VeVquench}, which shows a timescale of $\tau_R \, (L/\xi) \sim L/v$.

 \subsubsection{Equal time correlation functions}
 \label{sec:Gpipi}

 In mean-field theory, the pion and sigma correlators are obtained by linearizing the equations of motion around the time-dependent background from the previous section. For \( k\xi \ll 1 \) and \( t \sim \tau_R \), the coupling between the pion field \( \vec{\pi} \) and the conserved charge \( n_{ab} \) can be ignored. This follows from the full set of mean-field equations for the two-point functions in the time-dependent background (see \cref{app:meanfield}).  
More intuitively, due to charge conservation the charge and pion fields are coupled through a derivative interaction, which, over a limited time,  has little effect at small \( k \).  Initially, the equation of motion for \( \pi\pi \) matches that of \( \sigma\sigma \), since before the condensate forms, there is no preferred direction in flavor space and all field components must have the same correlation functions.
 
 At small momentum  and times $t \sim \tau_R$ 
 equations of motion for the equal time pion and sigma correlation
 functions read
 \begin{subequations}
  \label{eq:mean-field-pi-eom}
 \begin{align}
 \partial_t G_{\pi\pi} =&  -2 \Gamma_0  (m^2(t) +k^2)G_{\pi\pi} +2 \Gamma_0 \, ,  \\
 \partial_t G_{\sigma\sigma} =&  -2 \Gamma_0  (m^2_\sigma(t) +k^2)G_{\sigma\sigma} +2 \Gamma_0 \, ,
\end{align}
\end{subequations}
with the pion and sigma masses given by
\begin{subequations}
\begin{align}
    m^2(t)=& -|m_0^2| + \lambda\sigma^2(t)   \, ,  \\
    m^2_\sigma(t)=& -|m_0^2| + 3\lambda\sigma^2(t)  \, . 
\end{align}
\end{subequations}
In the limit when $k\xi \ll 1$,  we have $k^2 \ll |m_0^2| \sim \lambda \sigma^2(t)$ and  we can drop the $k^2$ in \labelcref{eq:mean-field-pi-eom} in this limit.  
Thus, in the linearized mean-field analysis the correlation functions become $k$ independent for $k\xi\ll 1$. 
This sharply contrasts with the scaling analysis in \cref{sec:scaling}, where the correlation functions scale as \( 1/(k\xi)^d \) in the same limit.

\begin{figure}
\includegraphics{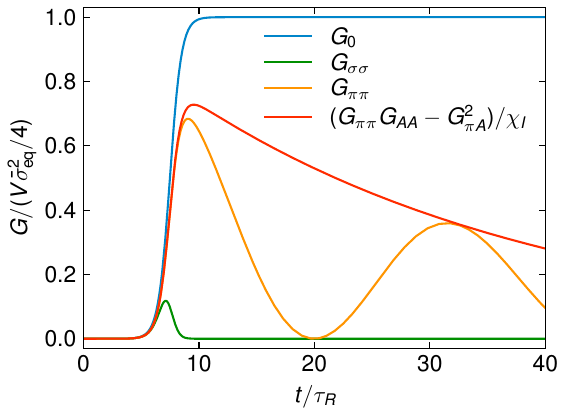}
\caption{Numerical results in mean-field theory ($d=4$) for the growth of the condensate, $G_0(t)$, and related equal time correlation functions $G_{\pi\pi}(t)$ and $G_{\sigma\sigma}(t)$ following a quench at time $t=0$ (see text).   The dimensionless parameters are $L/\xi=50$,   $D_0/\Gamma_0=1$,  $\lambda=1/4$,  and $v\tau_R/\xi=1$.  The condensate and correlators grow on a time $t_L \sim  \tau_R \, \ln((L/\xi)^{d/2})$ (see \cref{eq:t0def} for a definition) indicated in the plot. After a time $t_L$,  the pion waves with wave number $k=2\pi/L$ (shown here) oscillate on a time scale of $L/2v \sim \tau_R (L/\xi)$. 
}
\label{fig:meanfield}
\end{figure}

In the quench setup, the initial conditions for \( G_{\pi\pi}(t,k) \) and \( G_{\sigma\sigma}(t,k) \) are given by equilibrium at a reduced temperature \( \tredini \) (see \cref{eq:meanfieldsusc}). For \( k\xi \ll 1 \), both correlation functions start as \( G_{\pi\pi}(0,k) = G_{\sigma\sigma}(0,k) \simeq \chi \).  
In this limit the equations are solved using the approximation scheme from the previous section. At early times, when \( \lambda \sigma^2(t) \) is still small, the inhomogeneous term in \labelcref{eq:mean-field-pi-eom} (arising from thermal noise)  cannot be ignored. However, at late times, thermal noise becomes negligible while \( \lambda \sigma^2(t) \) plays a dominant role. By matching the late-time solution with the initial growth, as done in the previous section, we obtain the solution for \( G_{\pi\pi} \) and \( G_{\sigma\sigma} \) for  \( \tau_R \ll t \sim t_L \):
\begin{align}
 \frac{G_{\pi\pi}(t) }{\chi}
 &=  \frac{1}{4\lambda} \left(\frac{L}{\xi}  \right)^d \,   \frac{  e^{2(t-
 t_L)/\tau_R } }{1+e^{2(t - t_L)/\tau_R}  } , \label{eq:meanfieldGpipi}\\ 
         \frac{G_{\sigma \sigma}(t)}{\chi} &= \frac{1}{4\lambda} \left(\frac{L}{\xi} \right)^d
     \frac{  e^{2(t - t_L)/\tau_R } }{(1+e^{2(t
  - t_L)/\tau_R} )^3 } \, . \label{eq:meanfieldGsgsg}
\end{align}
The time evolution of $G_{\pi\pi}(t)/\chi$ exactly follows with $G_0/\chi$ from \labelcref{eq:meanfieldG0}. As anticipated, this implies that  $G_{\pi\pi}$ is amplified by the factor  $V\bsigma^2/4\chi$ and this enhancement  persists even at late times.  In contrast, $G_{\sigma\sigma}/\chi$ is also amplified but reaches a peak value of $4/27$ (in units of the amplification factor) before decreasing  exponentially. These analytic results hold for $k\xi\ll1$.

In \cref{fig:meanfield} we show numerical simulations for the correlation functions,  $G_0$, $G_{\pi\pi}$ and $G_{\sigma\sigma}$ for $L/\xi=50$ and $k = 2\pi/L$. 
As expected from \labelcref{eq:meanfieldG0,eq:meanfieldGpipi,eq:meanfieldGsgsg},
 the correlation functions grow to a size of order $V\bsigma^2/4$  over a timescale of $\sim \tau_R \log((L/\xi)^2)$. 
 At later times,  when $t \sim 1/vk$,  the coupling between the  pion $\vec{\pi}$ and  the conserved charges $n_{ab}$ can not be neglected and is analyzed \cref{app:meanfield}. At this stage $G_{\pi\pi}$ and the axial charge correlation function  $G_{AA}$  oscillate together with an angular frequency of $vk$ as exhibited in \cref{eq:oscillationsolution}.  The amplitude of the oscillations is determined by the
determinant $(G_{\pi\pi}G_{AA} - G_{A\pi}^2)/\chizero$ of the coupled system (see \cref{eq:oscillationdet} for the analytic expression), which gradually decays on the timescale for diffusive processes  $\sim 1/D_A k^2 \sim \tau_R /(k\xi)^2$.

In summary, the linearized mean-field approach accurately reproduces critical scaling and the expected parametric  size $\propto (L/\xi)^d$ of the enhancement in four dimensions (\( d = 4 \)). When a magnetic field is included as analyzed in \cref{app:meanfield}, the mean-field correlation functions are amplified by a factor of \( 1/m^4 \), consistent with the general scaling prediction of \( 1/(m\xi)^d \) only for \( d = 4 \). However,  in mean-field theory, the condensate grows exponentially over a timescale of \( \sim \tau_R \log((L/\xi)^2) \), rather than spreading ballistically over \( \sim \tau_R \, (L/\xi) \). Finally, the linearized mean-field approach fails spectacularly at capturing the \( k \)-dependence of the enhancement. In our statistical simulations the enhanced correlator scales as \( 1/(k\xi)^d \) for $k\xi\ll 1$, whereas in mean-field theory the correlator is enhanced, but independent of \( k \) until $k\xi \sim 1$.

 \section{Discussion}\label{sec:discussion}

Motivated by the rapid transition through the chiral cross-over in high-energy heavy-ion collisions,
we performed the first simulations of Model G, quenched from the high-temperature symmetric phase to the low-temperature broken phase in the chiral limit. 
We measured how the average chiral condensate and the 
equal-time correlation functions of the pion and  sigma modes approach equilibrium. In the quench setup, these curves are shown in \cref{fig:VeVquench,fig:quench,fig:sgquenchk}, along with a scaling analysis summarized below. 

The main result is a  parametric enhancement of the equal time pion-pion correlation function over equilibrium,  $G_{\pi\pi}(t, k)/G_{\pi\pi}^{\rm eq}(k)$ by a factor of order $\sim 1/(k\xi)$ (see \cref{fig:quench}). The enhancement sets in for $k\xi  \ll 1$ and  persists until late times, $t \sim \tau_R /(k\xi)$,  before slowly decaying  back to equilibrium in a time of order $\sim \tau_R/(k\xi)^2$. This is illustrated schematically in \cref{fig:Gpipisingle}.  

\DT{
With the quasiparticle assumptions outlined in \cref{sec:quasi},  the equal time two point function  parametrizes the  pion distribution function $n_\pi(t,k)$ at small $k$,  $\chi_I G_{\pi\pi}(t,k)/\bsigma^2 = n_{\pi}(t, k)/\omega(k)$ with $\omega(k)=vk$.  Our scaling results for $G_{\pi\pi}(t,k)$ can then be written\footnote{\DT{This expression is simply a rewrite of $G_{\pi\pi}(t,k)$ in \labelcref{eq:Gpipik} after using $\xi \propto v\tau_R$ and restoring the temperature.  It must be emphasized that the identification with the pion yield is only strictly valid for $vkt \gtrsim 1$ when the quasi-particle assumptions hold. } }
\st
n_{\pi}(t, k) = T_c \tau_R
\left[ \frac{1}{(vk \tau_R)^2} {\mathcal F}(v k t) \right] \, . 
\stp
The scaling function ${\mathcal F}(v k t)$ approaches  a constant for $ v k t\gg 1 $.  
For a given quench initial condition, $v$ and $\tau_R$ are fixed, and the spectrum exhibits a scaling form in time and momentum, which is characteristic of a Non-Thermal Fixed Point (NTFP)~\cite{Berges:2008wm,Berges:2014bba,Boguslavski:2018phs,Mikheev:2023juq}\footnote{\DT{
At a NTFP, the spectrum takes a non-equilibrium scaling form
$n(t, k) \propto t^{a} \mathcal{F}_S(t^{b} k)$
for some exponents $a$ and $b$. In the present case, $a = 2$ and $b = 1$. The number of soft pions in this non-equilibrium spectrum is not conserved, and the familiar NTFP relation $a = d b$ does not hold. }
}.
Alternatively, one may fix $k$ and vary the initial conditions. In this case, the non-equilibrium spectrum still shows a critical scaling with time and the correlation length $\xi$, reflecting the proximity to $T_c$. Both scalings are shown in \cref{fig:quench}.

Noting that the equilibrium spectrum is $n_\pi^{\rm eq} \simeq T_c/vk$ with $v\sim \xi/\tau_R$,   and taking the limit $v k t \gg 1$,   we can express the enhanced pion yield more precisely:
\st
n_{\pi}(t, k) = n_{\pi}^{\rm eq}(k) \left( \frac{\rm const}{k\xi} \right) \,  \qquad  \mbox{with} \qquad  \frac{\tau_R}{t} \ll k\xi  \ll   \left(\frac{\tau_R}{t} \right)^{1/2} \, . 
\stp
The constant is roughly $2.0$, as can be estimated from
the maximum of $G_{\pi\pi}(t,
k)/G_{\pi\pi}^{\rm eq}(k)$ shown in
\cref{fig:quench}. A more precise determination
will require larger lattices, where the NTFP
can be clearly separated from the approach to
equilibrium, and where the classical evolution
at high wavenumbers can be matched onto a
Boltzmann equation~\cite{Grossi:2020ezz}.
}

The parametric enhancement can be understood through a scaling analysis from \cref{sec:scaling},  along with a physical picture based on the hydrodynamic limits of Model $G$,  as illustrated in \cref{fig:drawing1}.  Specifically, over a time $\tau_R \propto \xi^{d/2}$,  disoriented domains of the chiral condensate form,  each  only a few correlation lengths long. The domains then  merge on a much longer time scale of order $t \sim L/v \sim \tau_R (L/\xi)$.  
\DT {Since the condensate growth timescale  is much larger than
    the critical  relaxation time $\tau_R$,  and, since the
    spatial scale $v t \sim L$ is much larger than $\xi$, the
    growth is determined by non-linear dynamics of the
    hydrodynamic limit of Model G. In the broken phase, this
    hydrodynamic limit corresponds to a $O(4)$ superfluid,
    first written down by Son in a rather different language\footnote{ \DT{In \cref{sec:pionoverview} we rederived Son's equations with an $O(4)$ notation (see \cref{eq:idealsuper}) making the connection with Model G transparent. }}~\cite{Son:1999pa}.
    The enhancement
    arises when an $O(4)$ superfluid is initialized with
    white noise and then subsequently evolves to form a global chiral condensate.  Because no direction in the $O(4)$ space is preferred, all directions try to become the global condensate, leading to a pion enhancement during the competitive growth. In the future it should be possible to simulate the superfluid limit directly without the full structure of Model G. 
}

We also investigated quench dynamics in mean-field theory, where the pion modes are treated as linearized fluctuations on top of a time dependent chiral condensate. Mean-field predicts scaling in four dimensions ($d=4$) and produces an enhancement of the right order of magnitude. However, the growth of both the enhancement and the chiral condensate is exponential rather than ballistic, saturating in a time of order $\tau_R \ln ((L/\xi)^2)$ instead of $\tau_R (L/\xi)$. Most importantly,  the linearized treatment gives a qualitatively incorrect dependence on $k\xi$. It remains to be seen whether taking into account a full non-linear treatment of mean field will give an improved prediction. We leave this discussion for future work.

An exciting phenomenological application of our result is the potential enhancement of soft-pion yields, which are currently underpredicted by hydrodynamic models that neglect critical dynamics. The difficulty in describing the experimental results with existing models is that the yield steadily rises above from the equilibrium curve only at small momenta. In current models there is no scale that would dictate this change.  While it is too early to draw a firm conclusion, our analysis naturally explains  why the yield of Goldstone pions should follow this trend.

\begin{figure}
  \centering
  \includegraphics[width=0.55\textwidth]{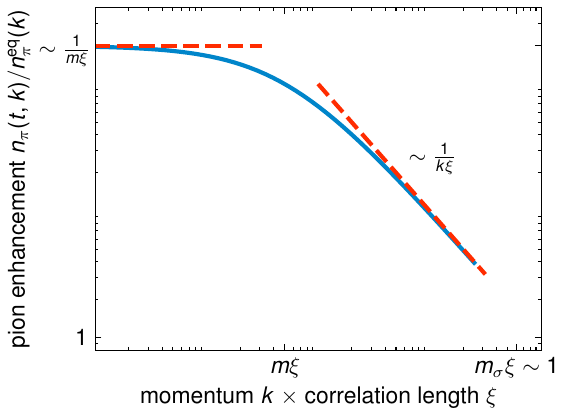}
  \caption{Schematic illustration of the expected pion enhancement as a function of momentum in units of QCD correlation length.}
  \label{fig:scaling_markup}
\end{figure}

Since our simulations were performed at zero magnetic field,  i.e., without explicit symmetry breaking,   an important question is how our results will be affected by finite pion mass $m$, or equivalently,  a finite magnetic field $H$. Including the pion mass is also essential for any comparisons with experiments. While a full study is left for future work, the scaling analysis of the quenches discussed in \cref{sec:scaling} can be straightforwardly extended to small but finite $m$.
In particular, the scaling function in \labelcref{eq:Gpipik} describing $G_{\pi\pi}$ for $k\xi \ll 1$  now  depends  on the dimensionless ratio,  $k/m$. 
In the regime $k \ll m \ll 1/\xi$,  the result should be independent of $k$,  leading to the prediction
 \begin{align}\label{eq:gpipi-predict}
     \left. G_{\pi\pi}(t, m) \right|_{t \sim 1/v m}  \sim  \xi^{2-\eta} \left(\frac{1}{m\xi}\right )^d  
 \end{align}
 This is an enhancement over equilibrium by a factor of order $\sim 1/(m\xi)$ which persists a long time,  
         $1/vm  \lesssim t \ll \tau_R/(m\xi)^2$.
 The result is intuitive: the $1/k\xi$ dependence is regulated by the pion mass at small momenta as  is illustrated in \cref{fig:scaling_markup}. 
 This conclusion is also supported by \DT{preliminary results of Langevin simulations at a finite magnetic field} and the mean-field analysis (see \cref{app:meanfield}), though in the latter case,  the enhancement is $1/(m\xi)^2$,  reflecting the fact that  mean-field is valid only in $d=4$.

\DT{
If real-world QCD lies outside the $O(4)$ scaling
window~\cite{Braun:2020ada,Braun:2023pew}, exhibiting only broken chiral
symmetry and a fairly light pseudo-Goldstone mode, the qualitative reasons for the enhanced pion-yield remain unchanged.   Critical scaling is essential for the data collapse seen in the top two panels of \cref{fig:quench}. However,  in the bottom two panels where the pion yield is also enhanced,  the initial conditions are fixed and the critical scaling of the parameters is never used. Thus,  outside of the scaling window  we expect that the superfluid dynamics will coarsen the system up to a length scale $m^{-1}$,  producing the enhancement as described previously. However,  since the superfluid parameters, $f^2$ and $m^2$, do not exhibit strict scaling, the data collapse with reduced temperature will not be perfect. In the language of the preceding paragraphs, outside of the  $O(4)$ scaling window we expect NTFP scaling but not critical scaling.
}

We have considered only instantaneous quenches. In a real heavy ion collision the QGP is expanding and cooling down at a finite rate. Therefore, it will be important to study different non-equilibrium protocols. Preliminary work where the reduced temperature is changed at a finite rate (as opposed to a sudden quench) shows that the basic picture of condensate growth outlined in \cref{fig:drawing1} remains valid~\cite{superpaper6}.  Indeed, this preliminary work motivated the current study of a theoretically cleaner case of instantaneous quenches.  

To summarize, we  highlight the following points from our work (see also the companion letter~\cite{Florio:2025zqv}):  
\begin{itemize}
    \item  We investigated the non-equilibrium dynamics of Model G, a universal model belonging to the same dynamic universality class as QCD.  
    \item  As the system passes through the phase transition, the equal-time correlation function of the Goldstone mode shows a parametric enhancement of order $1/(k\xi)$ compared to equilibrium  for $k\xi\ll 1$. \DT{In the quasi-particle picture (see \cref{sec:quasi}) this corresponds to enhancement of soft pion spectra, see \cref{fig:scaling_markup}.} 
 \item The non-equilibrium enhancement can be predicted using a general scaling argument (see \cref{sec:scaling}), which identifies three distinct timescales: relaxation, ballistic transport, and diffusion, as illustrated in \cref{fig:Gpipisingle}. 

 \item The growth of the chiral condensate and the enhancement of the pion yield takes place on timescales which are long compared to the dynamical relaxation time $\tau_R$. This is a hydrodynamic regime  and the appropriate non-linear effective theory is an ideal non-abelian superfluid, reflecting the broken $SU(2)_L\times SU(2)_R$ symmetry below $T_c$~\cite{Son:1999pa}. 
 \item The parameters of the hydrodynamic effective theory, which reflect the magnetic equation of state, can be determined  from Euclidean measurements in lattice QCD~\cite{Son:2002ci,Krasniqi:2024kwm}. This means that the details of the enhancement in QCD are rigorously computable  using current technology. 
\end{itemize}

\begin{acknowledgements}
      We would like to thank 
         Jürgen Berges, 
   Jean-Paul Blaizot, Kirill Boguslavski,
   Jannis Gebhard, 
   Alexander Kalweit,
   Govert Nijs,
   Thimo Preis, 
   Klaus Reygers, 
   Thomas Schäfer,
   Wilke van der Schee,
   Dam T. Son, 
   and Urs Wiedemann
      for helpful
     discussions.
     This work is supported by the DFG through Emmy Noether Programme (project number 545261797 (A.F.) and project number 496831614 (A.M.)) and CRC 1225 ISOQUANT (project number 27381115 (A.M.)). A.S. was supported by funding from Horizon Europe research and innovation programme under the Marie Skłodowska-Curie grant agreement No.~101103006 and the project N1-0245 of Slovenian Research Agency (ARIS). This work was supported by U.S. Department of Energy, Office of Science, Office of Nuclear Physics, Grant No. DE-SC0012704 (A.F.) and Grant
grant No. DE-FG-02-88ER40388 (D.T.).  This research used resources of the National Energy Research Scientific Computing Center (NERSC), a Department of Energy Office of Science User Facility using NERSC award NP-ERCAP0028307. We thank ECT* for support at the workshop ``Attractors and thermalization in nuclear collisions and cold quantum gases," where we profited from helpful discussions.
 \end{acknowledgements}
The data and plotting routines used in the figures are available at~\cite{florio_2025_17600429}.

 \appendix

\section{Subleading corrections to critical behavior\label{sec:subleading}}

Previous works~\cite{Florio:2021jlx,Florio:2023kmy} employed real-time simulations of Model G to study critical scaling in multiple dynamical channels. In particular, these studies extracted the nonuniversal leading and subleading amplitudes for several scaling observables. In the present work, whenever an equilibrium value is required we use the numerical values of these amplitudes together with the universal leading and subleading scaling exponents to compute the corresponding quantities. For convenience, we summarize the numerical values of the amplitudes in \cref{tab:subleading}.

Below $T_c$ the correlation length can be defined in terms of the pion decay constant $f$, see \cref{eq:corr_belowTc}. By analyzing the static correlator, we previously fitted $\bsigma^2 / f^2$ with a phenomenological ansatz $C_0 (-\tred)^{C_1}$~\cite{Florio:2023kmy}.\footnote{An incorrect value of $C_0$ was reported in Ref.~\cite{Florio:2023kmy}; the correct fit values are $C_0 = 0.923(8)$ and $C_1 = 0.009(3)$.}
Critical scaling predicts $\bsigma^2 / f^2 \sim (-\tred)^{\eta\nu}$ with a small exponent $\nu \eta \approx 0.02$ (see \cref{tab:num_values_fixed}). In practice, $\bsigma^2 / f^2 \simeq 0.9$ remains nearly constant over the temperature range relevant for our study. One could therefore use the condensate parametrization to determine the pion decay constant $f$ and the correlation length below $T_c$. However, we adopt a different strategy and use the definition of the pion velocity, \cref{eq:velocity}, to evaluate $f^2 = \chizero v^2$ and the correlation length below $T_c$.
We choose the normalization constant $\xi^-$ such that the correlation length exhibits a symmetric leading behavior around $T_c$, i.e.
$
\xi(\tred) = \xi^+ |\tred|^{-\nu},
$
where the nonuniversal amplitude $\xi^+$ above $T_c$ was determined in Ref.~\cite{Florio:2023kmy}. Note that with this definition, subleading corrections to the correlation length above $T_c$ and to the pion velocity below $T_c$ (see \cref{tab:subleading}) lead to a slight asymmetry of $\xi$ away from $\tred = 0$.

\begin{table}
  \centering
  \begin{tabular}{|c|c| c | } 
  \hline
  quantity & subleading scaling & constant values   \\
  \hline
any scaling quantity & $Y \propto \xi^{\theta_Y}(1+ C_Y \xi^{-\omega} + \ldots)$  &  $\omega=0.755(5)$ \\ 
susceptibility ($T>T_c)$ &  $\chi_+ = C^+ \tred^{-\gamma} (1+C_1 \tred^{\nu\omega})$ & $C^+ = 0.2077(32)$, $C_1=1.11(15)$   \\
  condensate ($T<T_c)$ &  $\bar{\sigma}_{\rm eq} = B^- (-\tred)^{\beta}(1+B_1 (-\tred)^{\nu\omega}) $
             & 
  $B^- = 0.988(7)$, $B_1=0.51(7)$ \\ 
  correlation length ($T>T_c$) & $\xi = \xi^+ \tred^{-\nu} (1+\xi_1 \tred^{\nu\omega})$ &   $\xi^+ = 0.4492(36)$, $\xi_1 = 0.332(85)$    \\
correlation length ($T<T_c$) & $\xi=  \xi^-/f^2 \equiv \xi^-/(v^2\chi_I)$ &  $\xi^-=\xi^+ (v^-)^2 \chi_I=0.479(7)$ \\ 
relaxation time ($T>T_c)$ & $\tau_R = \tau^+ \tred^{-\nu\zeta} (1+\tau_1 \tred^{\nu\omega})$  & $\tau^+=1.570(37)$, $\tau_1=-1.49(15)$ \\ 
pion velocity ($T<T_c$) & $v = v^- (-\tred)^{\nu/2} (1+v_1 (-\tred)^{\nu\omega})$  & $v^-=0.462(3)$, $v_1=0.325(3)$ \\ 
  \hline
\end{tabular}
\caption{The universal subleading scaling exponent $\omega$ and nonuniversal amplitudes of the $O(4)$ model at zero magnetic field used in this work. The first subleading exponent is taken from high-precision Monte Carlo simulations~\cite{Hasenbusch:2021rse}.  Nonuniversal leading and subleading amplitudes are taken from fits to Langevin simulations in Refs.~\cite{Florio:2021jlx,Florio:2023kmy}. 
}
\label{tab:subleading}
\end{table}
 
\section{
Details regarding the mean field theory
}\label{app:meanfield}

In this appendix, we will write down the model in mean field. Following the initial exposition, we will provide some further details on the quench from \cref{sec:mean-field-quench}. As in  \cref{sec:modelG} we will keep $g_0$ and $T_c$ explicit rather than setting them to unity as in the main text.  

The equations of the model are \cref{eq:eom1} and \cref{eq:eom2} with the Gaussian noises given in \labelcref{eq:langevin_var,eq:langevin_var_cons}.
It is useful to decompose the field and the charge into an orthonormal basis in flavor space as
\st
\phi_a(t, {\bm x}) \equiv   \sigma(t, {\bm x})  \, \nhat_a + 
\vec{\pi}(t, {\bm x}) \cdot \vec{e}_a\, ,
\stp
where $\vec{e}_a$ and $\hat n_a $ form an orthonormal basis in flavour space. 
Here, we derive the equation in the presence of the explicit breaking $H$. Therefore, it is more natural to have a basis fixed in time with the $\hat n$ component  directed along the explicit breaking,  $H_a =H \hat n_a $.
The charges are in the adjoint of the group, leading to the following decomposition
\st
n_{ab}(t, {\bm x})  =
\vec{n}_{A}(t, {\bm x})  \cdot \left( \nhat_a \, \vec{e}_b -  \nhat_b \, \vec{e}_a \right) 
+ 
\vec{n}_{V}(t, {\bm x})  \cdot  \left(\vec{e}_a \times \vec{e}_b \right).
\stp
We can then project the equation of motions into this basis:
\begin{subequations}
\begin{align}
   \partial_t \sigma  + \frac{g_0}{\chizero}\,\vec{n}_A \cdot \vec{\pi}  
  &=\Gamma_0 \nabla^2 \sigma - \Gamma_0 (m_0^2 + \lambda \sigma^2 + \lambda\vec{\pi}\cdot \vec{\pi})\sigma +
  \Gamma_0 H + \theta\, , \label{eq:sigmaequation} \\
   \partial_t \vec{n}_{V}  + g_0 \,\nabla \cdot (\nabla \vec{\pi} \times \vec{\pi }) 
             &= D_0 \nabla^2 \vec{n}_{V}+  \partial_{i} \vec{\Xi}_{V}^i\label{vector-equation},\\
      \partial_t \vec{\pi}  - \frac{g_0}{\chizero}\,\vec{n}_A \sigma 
  &=\Gamma_0 \nabla^2 \vec{\pi} - \Gamma_0 (m_0^2 + \lambda \sigma^2 + \lambda\vec{\pi}\cdot \vec{\pi})\vec{\pi}  + \vec{\theta},\label{eq:pion}\\
  \partial_t \vec{n}_A  + g_0 \,\nabla \cdot (\nabla \sigma \vec{\pi} -   \sigma \nabla\vec{\pi}) + g_0 H \vec{\pi} 
             &= D_0 \nabla^2 \vec{n}_A +  \partial_{i} \vec{\Xi}_{A}^i \label{eq:axial}.
\end{align}
\end{subequations}
The non-vanishing noise variances are: 
\begin{subequations}
\begin{align}
   \langle \theta(t,x)\theta(t',x') \rangle &= 2 T_c\Gamma_0 \, 
   \, \delta(t-t')\delta^3(x-x') \, ,\\
   \langle \theta^{\ell_1}(t,x)\theta^{\ell_2}(t',x') \rangle &= 2 T_c\Gamma_0 \, \delta^{\ell_1 \ell_2}
   \, \delta(t-t')\delta^3(x-x') \, ,\\
   \langle \Xi^{\ell_1i}_{V}(t,x)\Xi^{\ell_2 j}_{cd}(t',x') \rangle &= 2  T_c\chizero D_0 \,
   \delta^{ij} \delta^{\ell_1 \ell_2}
   \, \delta(t-t')\delta^3(x-x') ,
   \\
   \langle \Xi^{\ell_1i}_A(t,x)\Xi^{\ell_2j}_{A}(t',x') \rangle &= 2  T_c \chizero D_0 \,
   \delta^{ij} \delta^{\ell_1\ell_2} 
   \, \delta(t-t')\delta^3(x-x') . 
\end{align}
\end{subequations}

Following the spirit of mean field, we will linearize the equations of motion around a time dependent and spatially uniform condensate,  $\phi_a(t) = (\sigma(t), \vec{0})$. 
The evolution of the order condensate is non-linear and given by
\begin{align}\label{eq:vev-evol}
\partial_t \sigma   
   &=\Gamma_0\nabla^2 \sigma - \Gamma_0 (m_0^2 + \lambda \sigma^2)\sigma + \Gamma_0H,
\end{align}
which follows by substituting $\phi_a(t)=\sigma(t)n_a$ into \labelcref{eq:sigmaequation} and averaging over volume. For simplicity,   we have dropped the noise in the time evolution of the condensate since the averaged noise is suppressed by volume (see \cref{sec:volume} where the noise is included). 
Given this equation for the condensate, we can write down  the linearized equations for the spatially inhomogeneous fluctuations, $\delta\phi_a = (\delta \sigma, \vec{\pi})$
\begin{subequations}
    \label{eq:linall}
\begin{align}
   \partial_t \delta \sigma  
  &=\Gamma_0 \nabla^2 \delta\sigma - \Gamma_0 m_\sigma^2 \delta \sigma + \theta\, , \label{eq:lin1}\\
   \partial_t \vec{n}_{V} 
             &= D_0 \nabla^2 \vec{n}_{V}+  \partial_{i} \vec{\Xi}_{V}^i\, ,\label{eq:lin2} \\
      \partial_t  \vec{\pi}  - \frac{g_0}{\chizero}\, \vec{n}_A \sigma \, ,   
  &=\Gamma_0 \nabla^2  \vec{\pi} - \Gamma_0 m^2  \vec{\pi}  + \vec{\theta} \, , \label{eq:lin3} \\
  \partial_t \vec{n}_A  + g_0 \,\nabla \cdot (\nabla \sigma \vec{\pi} -   \sigma \nabla \vec{\pi}) + g_0 H  \vec{\pi} 
             &= D_0 \nabla^2  \vec{n}_A +  \partial_{i} \vec{\Xi}_{A}^i \, .\label{eq:lin4}
\end{align}
\end{subequations}
In writing these equations we adopted the following standard definition
of the mean-field masses
 \begin{equation}\label{eq:masses}
     m_\sigma^2(\sigma)\equiv m_0^2 + 3 \lambda \sigma^2,\quad m^2(\sigma) \equiv  
     m_0^2 +  \lambda \sigma^2 .
 \end{equation}

We define the (connected) correlators as 
\begin{subequations}
\begin{align}
G_{\sigma\sigma}(t,x,y) &= \langle
\delta \sigma(t, 
x )\delta \sigma (t,y)\rangle , \\
G^{\ell_1,\ell_2}_{\pi\pi}(t,x,y)&= \langle
 \pi^{\ell_1}(t, 
x ) \pi^{\ell_2} (t,y)\rangle=\delta^{\ell_1 \ell_2 }G_{\pi\pi}(t,x,y),  \\
G^{\ell_1,\ell_2}_{AA}(t,x,y)&= \langle
 n_A^{\ell_1}(t, 
x )n_A^{\ell_2} (t,y)\rangle=\delta^{\ell_1 \ell_2 }G_{AA}(t,x,y), \\
G^{\ell_1,\ell_2}_{A\pi}(t,x,y)&= \langle
 n_A^{\ell_1}(t, 
x ) \pi^{\ell_2} (t,y)\rangle=\delta^{\ell_1 \ell_2 }G_{A\pi}(t,x,y), \\
G^{\ell_1,\ell_2}_{VV}(t,x,y)&= \langle
 n_V^{\ell_1}(t, 
x ) n_V^{\ell_2} (t,y)\rangle = \delta^{\ell_1 \ell_2 }G_{VV}(t,x,y).
\end{align}
\end{subequations}
Here we have neglected non-Gaussian correlations in these definitions. Therefore, the isospin dynamics are trivial, as each correlator is diagonal in flavor space.

Given this definition of equal time correlation functions, the evolution equations can be deduced from the linearized equations in \labelcref{eq:linall}.
At present, we will outline the essential steps only for the scalar channel to focus the discussion.

For a short time step $\Delta t$, we have 
\begin{equation}
     \delta \sigma(t+\Delta t)  
  =\delta \sigma(t) +\Delta t\left( \Gamma_0 \nabla^2 \delta\sigma - \Gamma_0 m_\sigma^2 \, \delta \sigma\right) +  \Delta \theta\, , 
\end{equation}
where the Gaussian random variable $\Delta \theta$ is
\begin{equation}
  \Delta \theta = \int_{t}^{t+\Delta t } \mathrm d t^{\prime} \theta(t^{\prime})   . 
\end{equation}
With this definition, expanding in series of $\Delta t $, using the variance of the noise $\theta$, we find 
\begin{align}
    \partial_t G_{\sigma \sigma}(t, x,y ) 
 =&\Gamma_0 \left(\nabla_x^2 + \nabla_y^2\right) G_{\sigma \sigma}(t, x,y)
-2\Gamma_0 m_\sigma^2 G_{\sigma \sigma}(t,x,y) +2  \Gamma_0 T_c\delta^{3}(x-y).
\end{align}
In the same way, all the correlators can be worked out. It is more natural to express their dynamics in Fourier space,  yielding
\begin{align}
\partial_t G_{\sigma \sigma}(t, k ) 
& =-2\Gamma_0 ( k^2+ m_\sigma^2)\, G_{\sigma \sigma} (t, k) +2  \Gamma_0 T_c \, .
\end{align}
The vector channel correlator equation follows from \labelcref{eq:lin2}
\begin{align}
\partial_t G_{VV}(t, k)
   &=-2D_0k^2 G_{VV}  +2  D_0 k^2 \, T_c \chizero \, . 
\end{align}
The pion-axial correlators are coupled with the pion correlator given by 
\begin{subequations}
    \label{eq:pionaxial}
\begin{align}
   \partial_tG_{\pi\pi}(t,k) &= 
 \frac{g_0\sigma }{\chizero}\, \left( G_{A\pi}  +  G_{\pi A} \right)   
  -2\Gamma_0(k^2+m^2  )G_{\pi\pi}
  +2 \Gamma_0 T_c \, , 
  \label{eq:pipi}
\end{align}
the axial correlator  given by
\begin{align}
 \partial_tG_{AA}(t, k) =  
-g_0 (k^2\sigma  +H ) \, (G_{A\pi}  +  G_{\pi A})  - 2D_0k^2 G_{AA}
+2  D_0 k^2 \, T_c\chizero \, ,  
\end{align}
and,  finally,  the mixed  correlator given by
\begin{align}
\partial_tG_{A\pi}(t, k) =
-g_0 (\sigma k^2+H )  G_{\pi\pi} + \frac{g_0 \sigma}{\chizero}\, G_{AA}  - \Gamma_0(k^2+m^2) G_{A\pi}
 -D_0 k^2 G_{A\pi} .\label{eq:Api}
\end{align}
\end{subequations}
In the following, we will set $g_0=T_c=1.$

\subsubsection{Mean field description of a quench}

Having presented the mean-field equations, we will now describe the evolution of $G_{\pi\pi}(t, k)$ and $G_{AA}(t,k)$ following a quench for $k\xi \ll 1$. This expands on the discussion given \cref{sec:Gpipi}
.  For simplicity, the magnetic field $H$ will be set to zero in this section. 

As described in the body of the text, after the quench, when the condensate is on ``top of the hill'', the initial behavior of $G_{\pi\pi}$ is characterized by exponential growth \DT{(see \cref{fig:Gpipisingle})}.  The resulting evolution of the condensate and $G_{\pi\pi}$ are given in \labelcref{eq:sigmasol,eq:meanfieldG0,eq:meanfieldGpipi}.

During this short period of exponential growth, which lasts a time  of order $t_L \sim \tau_R \, \log((L/\xi)^2)$,  the axial correlators $G_{A\pi}$ and $G_{AA}$ are constant, taking values zero and $\chizero$, respectively. These initial values are the equilibrium values of these correlators before the quench  at temperature $\tredini$. These correlators are nearly constant because the axial charge is conserved and thus evolves slowly for $k\xi \ll 1$.  For the same reason,  the coupling between $G_{AA}$ and $G_{\pi\pi}$ can be neglected for times of order $t_L$. However,  at later times  $t \sim 1/(vk) \sim \tau_R /k\xi$,  the condensate has reached its equilibrium value and the coupling between the pion and axial fields can not be ignored. Here, we will characterize the late time oscillatory behavior of the correlators, which is seen in \cref{fig:meanfield} and reflects this coupling.

 At late times $t-t_L \gg \tau_R$,  the condensate has reached its asymptotic value of $\sigma = \bsigma$ and is treated as constant in what follows. Then we can write the evolution equations of the correlators in the pion-axial channel via $\vec{G}=(G_{\pi\pi},G_{A\pi},G_{\pi A},G_{AA})$ as 
 \begin{align}\label{mixed-sector-eqs}
 \partial_t {G}_i= A_{ij}G_j+s_i, 
 \end{align}
 where the sources are $s_i=(2\Gamma_0,0,0,2 D_0\chizero k^2)$ and the matrix is
\begin{align}
 A_{ij}=\begin{pmatrix}
 -2\Gamma_0 k^2 & \bsigma/\chizero & \bsigma/\chizero & 0\\
 -\bsigma k^2  & -( D_0 +\Gamma_0 )k^2 &0 & \bsigma/\chizero\\
 -\bsigma k^2  & 0  & - (D_0 +\Gamma_0) k^2& \bsigma/\chizero\\
 0&-\bsigma k^2  & -\bsigma k^2 & - 2D_0k^2
 \end{pmatrix},
 \end{align}
where we have set $H=0$.

We can easily determine the eigenvalues of the above matrix 
\begin{align}
    \lambda_\pm &= -D_A k^2  \pm\sqrt{(\Gamma_0 - D_0)^2k^4-4k^2 v^2},\label{eval2}
\end{align}
where we recalled the pion velocity $v^2\equiv \sigma^2/\chizero$, see \cref{eq:velocity},  and the damping coefficient $D_A \equiv \Gamma_0 + D_0$.
Sticking to small momenta, it is easy to see that the eigenfrequencies are given by
\begin{align}
    \omega(k) = i \lambda_\pm \sim \pm 2 v  k - iD_Ak^2 +\mathcal{O}(k^3),
\end{align}
in line with the discussion in \cref{sec:hydroGM}.

In the limit of $\Gamma_0=D_0=D_A/2$, these equations can be solved once the mean field reaches its constant equilibrium value at late times. The initial evolution for $t \sim t_L$ gives the initial conditions for the subsequent evolution with $t \sim 1/vk $.  This initial condition can be well approximated by 
\st
G_{\pi\pi}(0)  = \frac{V\bsigma^2}{4} \, , 
\stp
while the remaining components are much smaller $G_{AA} = \chi$ and $G_{A\pi}=0$. The solutions take the form of a homogeneous solution, which depends on $G_{\pi\pi}(0)$ and is initially large, and  an inhomogeneous solution, which is small and determines the late time  equilibrium behavior.
The solution takes the form
\begin{subequations}
    \label{eq:oscillationsolution}
\begin{align}
    k^2 G_{\pi\pi}&=1 + G_{\pi\pi}(0)k^2 \, e^{-D_A k^2 t} \, \cos^2(v k t) \, ,   \\
    G_{AA}/\chizero &= 1 + 
 G_{\pi\pi}(0)k^2\,  e^{-D_A k^2 t} \, \sin^2(vk t)  \, ,  \\ 
    kG_{\pi A}/\sqrt{\chizero}&= - G_{\pi\pi}(0)k^2  \, e^{-D_A k^2 t} \, \sin(vkt) \cos(vk t) \, .
\end{align}
\end{subequations}
In  this solution  the leading constants reflect the late time equilibrium behavior, while the  remaining terms proportional to $G_{\pi\pi}(0)$ represent the non-equilibrium  homogeneous solution.  In writing this solution we have made the approximation, $G_{\pi\pi}(0)k^2  - 1 \simeq  G_{\pi\pi}(0) k^2$, which is uniformly valid  at all times for $\xi/L \ll 1$.  In the same approximation,  \labelcref{eq:oscillationsolution} is valid even for $\Gamma_0\neq D_0$,  and the determinant depends only on the axial diffusion constant and not on the pion velocity:
\begin{equation}
    \label{eq:oscillationdet}
    \frac{k^2}{\chi_I} (G_{\pi\pi} G_{AA}-G_{\pi A }^2) = 
    1+G_{\pi\pi}(0)k^2 e^{-D_A k^2 t} 
    .
\end{equation}

For times of order $t \sim 1/vk$ the damping term $e^{-D_Ak^2 t}\simeq 1$ is approximately unity and the homogeneous terms proportional to $G_{\pi\pi}(0)$  in \labelcref{eq:oscillationsolution,eq:oscillationdet} are dominant. 
These terms account for qualitative and quantitative features of the late time behavior of the curves in \cref{fig:meanfield}.

 \subsection{Mean field at finite and small $H$ }
 \label{app:meanquench}

 \begin{figure}
     \centering
          \includegraphics[width=0.33\textwidth]{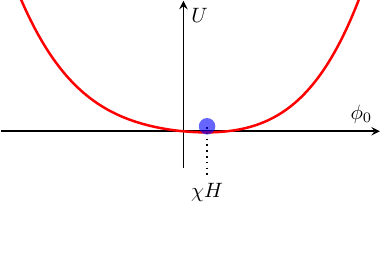}%
     \includegraphics[width=0.33\textwidth]{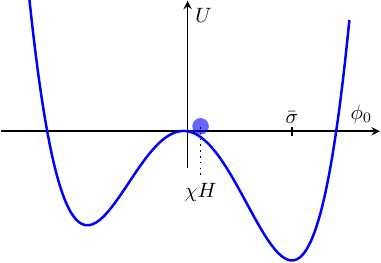}%
     \includegraphics[width=0.33\textwidth]{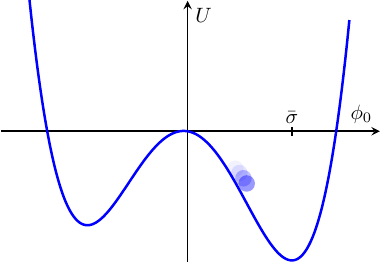}
     \caption{ Schematics of quenching from high-temperature to the broken phase
     in the presence of explicit symmetry breaking $H\neq 0$. The order parameter is initiated in thermal equilibrium. The finite
 magnetic field introduces a displacement $\chi H$. The instantaneous quench
 brings the potential to its broken phase. The fields roll down to the bottom of one of the wells.   }
     \label{fig:drawing2}
 \end{figure}
 \subsubsection{Overview}
 In \cref{sec:volume} we analyzed the growth of the condensate following a quench seeded by an initial fluctuation $M_a \sim  \sqrt{\chi/V}$.  In this section,  we will modify this analysis by considering a small magnetic field which provided a small initial value  which seeds the growth of the condensate.   The  analysis of the subsequent growth of the condensate and the correlation functions $G_{\pi\pi}$ and $G_{\sigma\sigma}$ is quite similar to \cref{sec:volume}.

 Consider the field configuration pre-quench shown in the left panel of \cref{fig:drawing2}. 
When a small magnetic field is applied, the mean value of the zero mode is non-zero even above $T_c$ and equals
 \st
 \label{eq:sigma0-initial}
   \llangle M_0 \rrangle = \sigma(0)  = \chi H  \,.
 \stp
Here the mean-field the susceptibility is given in \cref{eq:meanfieldchi}.  Subsequently
$M_0$ grows exponentially from this initial size to  $\bsigma$ as the condensate evolves from the unstable configuration in the left panel of \cref{fig:drawing2} to the stable one in the rightmost plot of \cref{fig:drawing2}. The  process leads to amplification factor for $M^2$ of order 
\st
\label{eq:Hamplification}
{\rm amplification}  =  \frac{\bsigma^2}{(\chi H)^2}  =  \frac{1}{(m_{H} \xi)^d} \, . 
\stp
Here we recalled the definition of $\chi=\xi^2$ above $T_c$ given in \cref{eq:meanfieldchi}  
and also defined the equilibrium pion screening mass
\st
m^2_{H}  \equiv  \frac{H}{\bsigma} \, , 
\stp
which follows from the form of the static two point functions of the order parameter $\delta \phi_a = (\delta \sigma(k), \vec{\pi}({\bm k}) )$ in equilibrium    below $T_c$  and in mean-field~\cite{Grossi:2021gqi}.
This definition agrees with \cref{eq:masses} only when the condensate has reached its equilibrium value of $\bsigma$.  Finally, as in the mean field results of \cref{sec:volume},  we have assume $d=4$ when presenting the mean field amplification in \cref{eq:Hamplification}. Note that we introduce a subscript $H$ to the mass to distinguish from the rest of the discussion in the main text with the exception of the mass in \eqref{eq:gpipi-predict}, which is $m_H$.

The timescale for exponential growth is $\tau_R$, and  it takes $\bsigma/\chi H$  e-foldings for the condensate to reach its asymptotic value.  Thus, in analogy with \cref{sec:volume}  we will define a timescale
\st
\label{eq:tH}
t_H \equiv \tau_R \ln\left(\frac{\bsigma}{2\chi H} \right) = \tau_R \ln\left(\frac{1}{2 (m_{H} \xi)^{d/2}}\right) \, , 
\stp
where the additional factor of $2$ which appears in the logarithm is conventional following from the computations given below.

 \subsubsection{Growth of the condensate and correlations}
 Having outlined the timescales and scaling in the previous section, we will now present the details.
 The dynamics of condensate growth proceeds in two stages, paralleling   \cref{sec:volume}. Initially, $\sigma$ is of
 the order of the perturbation provided by $H$. Non-linearities can be ignored, and $\sigma$ grows exponentially. In the second stage, when $\sigma/\chi H \gg 1$, the
 perturbation $H$ is irrelevant, but the dynamics become non-linear.

 Let us see this explicitly. 
 The evolution of $\sigma$ can be determined from
 \cref{eq:vev-evol} by
 \st
 \label{eq:dt_meanfield}
  \partial_t  \sigma = -\Gamma_0 \left( m^2(\sigma) 
  \sigma - H   \right),
 \stp
 where the transverse mass is given by \cref{eq:masses}.
 When
 $\lambda \sigma^3 \ll \sigma m_0^2$,  we can neglect the
 non-linearities\footnote{Note that this regime $\lambda \sigma^3 \ll
 \sigma m_0^2$ is equivalent to $\sigma \ll \sigma_{\rm eq}$.} and obtain
 \st 
    \partial_t \sigma =  \Gamma_0 m_0^2\sigma  + \Gamma_0 H .
 \stp
 The solution in this limit reads
 \st
 \sigma(\tau) = \sigma(0) e^{t/\tau_R} +  \chi H e^{t/\tau_R} (1 - e^{-t/\tau_R}).
 \stp
 Upon using the initial condition \cref{eq:sigma0-initial}, we see that the condensate grows exponentially for $t \gg \tau_R$
 \st
 \label{eq:sigmaoftshortH}
  \sigma(t) = 2\chi H  e^{t/\tau_R}\,  . 
 \stp
Ultimately,   $\sigma$ cannot be considered small anymore and the non-linearities becoming relevant.

 After this initial stage, however,  the magnetic field is small compared to the
 field $\sigma \gg \chi H$, and the magnetic field can be neglected. 
 In the case of vanishing magnetic field, the solution of the nonlinear equation
 \cref{eq:dt_meanfield} with $H=0$ is 
 \begin{align}
     \label{eq:dt_sigma-H0}
                  \sigma^2(t)&=\frac{ \sigmaeq^2}{e^{ -2  (t- t_H)/\tau_R 
                  }+1}, 
 \end{align}
 where at this stage the timescale $t_H$ is an integration constant. 
 The integration constant has to be determined by the early time behavior in \labelcref{eq:sigmaoftshortH} to \labelcref{eq:dt_sigma-H0} yielding
 \st
 e^{t_H/\tau_R} =   \frac{\bsigma}{2\chi H},
 \stp
 which agrees with the previously definition of $t_H$ given in \labelcref{eq:tH}

Summarizing,  following a quench, the condensate squared  (relative to its initial  value) grows as 
\st
G_0(t) \equiv \left(\frac{\sigma(t)}{\chi H}\right)^2  = \frac{1}{(m_H \xi)^d }  \,  \frac{ e^{2(t - t_H)/\tau_R }}{1 + e^{2(t - t_H)/\tau_R }  }  \, , 
\stp
increasing  by a factor of order  $1/(m_H \xi)^d$ over a timescale of order $t_H \sim \tau_R \ln \left((m_H \xi)^{-d/2}\right)$, c.f. \cref{eq:Hamplification,eq:tH}.
 
\subsubsection{Equal time correlation functions}

The pion and sigma correlators can be determined following a similar strategy.   
For $k^2 \ll m_0^2$ and the approximations discussed in the body of the text (\cref{sec:volume}),  the correlators evolve as
\begin{align}
    \partial_t G_{\pi\pi} =&  -2 \mGamma  m^2(t) G_{\sigma\sigma}   + 2 \mGamma  \, ,  \\
    \partial_t G_{\sigma\sigma} =&  -2 \mGamma  m_\sigma^2(t) G_{\sigma\sigma}   + 2 \mGamma  \, ,
\end{align}
where the time dependence of the masses stems from the time dependent condensate, e.g. $m^2(t) = -|m_0^2| + \lambda \sigma^2(t)$. 
These equations are subject to the initial condition 
\st \label{eq:gsigini}
G_{\sigma\sigma}(0)  = G_{\pi\pi}(0)= \chi \, .  
\stp
which reflects the equilibrium at $\tredini$ just before the quench. 

As previously, at early times, when $\lambda \sigma^2(t)$ is small compared to $-|m_0|^2$, the inhomogeneous terms (arising from thermal noise) cannot be ignored. However, at late times the inhomogeneous term becomes negligible, while $\lambda \sigma^2$ plays a dominant role. By matching the initial growth with the late time solution with  the initial growth we obtain 
\begin{align}
 \frac{G_{\pi\pi}(t) }{\chi}
 &=  \frac{1}{2} \left(\frac{1}{m_H \xi}  \right)^d\,   \frac{  e^{2(t-
 t_H)/\tau_R } }{1+e^{2(t - t_H)/\tau_R}  } , \label{eq:meanfieldHGpipi}\\ 
         \frac{G_{\sigma \sigma}(t)}{\chi} &= \frac{1}{2} \left(\frac{1}{m_H \xi} \right)^d
     \frac{  e^{2(t - t_H)/\tau_R } }{(1+e^{2(t
  - t_H)/\tau_R} )^3 } \, . \label{eq:meanfieldHGsgsg}
\end{align}
\
\

\begin{figure}
\includegraphics{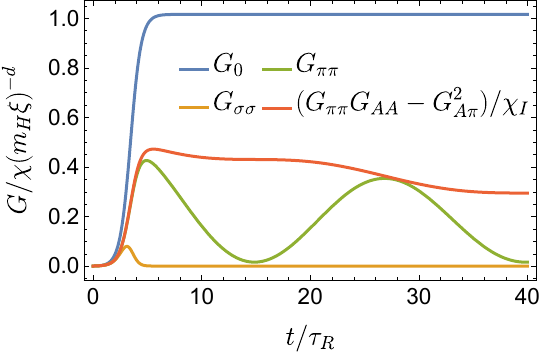}
\caption{Numerical results in mean-field theory ($d=4$) for the growth of the condensate, $G_0(t)$, and related equal time correlation functions $G_{\pi\pi}(t)$ and $G_{\sigma\sigma}(t)$  following a quench at time $t=0$ for $k=0$ and  for finite magnetic field, $m_H\xi = 2\pi/50$ (see \cref{fig:meanfield} for comparison).  The timescale for condensate growth is $t_H \equiv \tau_R \ln((m_H \xi)^{-d/2}/2) \simeq 4.84\,\tau_R$.   The remaining dimensionless parameters are  $D_0/\Gamma_0=1$,  $\lambda=1/4$,  and $v\tau_R/\xi=1$.  
}\label{fig:meanfieldfiniteH}
\end{figure}

In writing these equations we noted that the amplification factor $(\bsigma/\chi H)^2$ can be written as $(m_H\xi)^{-d}$ via \cref{eq:Hamplification}. These finite $H$ results are similar to the finite volume results, see \cref{eq:meanfieldGpipi,eq:meanfieldGsgsg}. Moreover, the correlators at finite $H$ will look similar to those in \cref{fig:meanfield}, with the substitution
\st
    \frac{1}{4\lambda} \left(\frac{L}{\xi}\right)^d\rightarrow (m_H
    \xi)^{-d}.
    \stp
     Omitting the details,  at late times the axial charge and
    pion correlators will oscillate together with frequency $v m_H$ and decay
    rate $\Gamma_0 m^2_H$,  instead of $vk$ and $D_A k^2$  as discussed in
    \cref{eq:oscillationsolution}. A numerical solution of the mean field
    equations is shown in \cref{fig:meanfieldfiniteH} and confirms this
    analysis. 

\bibliography{literature}

\end{document}